\documentclass[useAMS,usenatbib]{mnras}
\usepackage{graphicx,natbib,amssymb,amsmath,stmaryrd,makecell}
\usepackage{txfonts}
\usepackage{xcolor}
\usepackage{hyperref}
\usepackage{xspace}
\usepackage{caption}
\usepackage{lipsum}
\usepackage{soul}

\newcommand{\GHz}{{\rm GHz}}

\newcommand{\expf}[1]{{{\rm e}^{#1}}}

\newcommand{\TCMB}{T_{\rm CMB}}

\newcommand{\Planck}{{\it Planck}\xspace}

\newcommand{\id}{{\,\rm d}}

\newcommand{\beq}{\begin{equation}}   %

\newcommand{\eeq}{\end{equation}}   %

\newcommand{\beqa}{\begin{eqnarray}}   %

\newcommand{\eeqa}{\end{eqnarray}}   %

\newcommand{\beal}{\begin{align}}

\newcommand{\enal}{\end{align}}

\newcommand{\bspl}{\begin{split}}

\newcommand{\espl}{\end{split}}

\newcommand{\bsub}{\begin{subequations}}

\newcommand{\esub}{\end{subequations}}

\newcommand{\bmulti}{\begin{multline}}   %

\newcommand{\beqm}{\begin{mathletters}}   %

\newcommand{\eeqm}{\end{mathletters}}   %

\newcommand{\Te}{T_{\rm e}}

\newcommand{\vek} [1]{\mbox{\boldmath${#1}$\unboldmath}}

\newcommand{\pot}[2]{#1 \times 10^{#2}}

\newcommand{\COBEF}{{\it COBE/FIRAS}\xspace}

\newcommand{\PIXIE}{{\it PIXIE}\xspace}

\newcommand{\Tin}{T_{\rm in}}
\newcommand{\zBBN}{z_{\rm BBN}}
\newcommand{\zcon}{z_{\rm con}}
\newcommand{\xin}{x_{\rm in}}
\newcommand{\eV}{{\rm eV}}
\newcommand{\keV}{{\rm keV}}
\newcommand{\mdp}{m_{\rm d}}

\newcommand{\gammacon}{\gamma_{\rm con}}

\newcommand{\mucon}{\mu_{\rm con}}

\author[Chluba, Cyr, and Johnson]{ 
Jens Chluba$^1$\thanks{E-mail:jens.chluba@manchester.ac.uk},
Bryce Cyr$^{\,2,1}$\thanks{E-mail:brycecyr@mit.edu},
and Matthew C. Johnson$^{3,4}$
\\
$^1$Jodrell Bank Centre for Astrophysics, School of Physics and Astronomy, The University of Manchester, Manchester M13 9PL, U.K.
\\
$^2$Center for Theoretical Physics, Massachusetts Institute of Technology, Cambridge, MA 02139, U.S.A.
\\
$^3$Perimeter Institute for Theoretical Physics, 31 Caroline St N, Waterloo, ON N2L 2Y5, Canada
\\
$^4$Department of Physics and Astronomy, York University, Toronto, ON M3J 1P3, Canada
\vspace{-2mm}
}

\date{\vspace{-3mm}Accepted XXX. Received YYY; in original form ZZZ}

\title[Dark Photon Spectral Distortions]
{Revisiting Dark Photon Constraints from CMB Spectral Distortions}

\begin{document}

\maketitle

\begin{abstract}
Spectral distortions of the cosmic microwave background (CMB) provide stringent constraints on energy and entropy production in the post-BBN (Big Bang Nucleosynthesis) era. This has been used to constrain dark photon models with \COBEF and forecast the potential gains with future CMB spectrometers. Here, we revisit these constraints by carefully considering the photon to dark photon conversion process and evolution of the distortion signal. Previous works only included the effect of CMB energy density changes but neglected the change to the photon number density. We clearly define the dark photon distortion signal and show that in contrast to previous analytic estimates the distortion has an opposite sign and a $\simeq 1.5$ times larger amplitude. We furthermore extend the treatment into the large distortion regime to also cover the redshift range $\simeq \pot{2}{6}-\pot{4}{7}$ between the $\mu$-era and the end of BBN using {\tt CosmoTherm}. This shows that the CMB distortion constraints for dark photon masses in the range $10^{-4}\,\eV\lesssim \mdp\lesssim 10^{-3}\,\eV$ were significantly underestimated. We demonstrate that in the small distortion regime the distortion caused by photon to dark photon conversion is extremely close to a $\mu$-type distortion independent of the conversion redshift. This opens the possibility to study dark photon models using CMB distortion anisotropies and the correlations with CMB temperature anisotropies as we highlight here.
\\
\end{abstract}

\section{\label{sec:level1}Introduction}
The cosmic microwave background (CMB) contains an enormous wealth of information, and has played a key role in ushering in the current era of precision cosmology. Dedicated studies of CMB temperature anisotropies over the past decades has culminated in the $\Lambda$CDM model \citep{Planck2018params}, which describes much of the observable Universe through a model with only seven independent parameters. 
Complementary to the temperature anisotropies, the frequency spectrum of CMB photons is capable of providing deep insights into the thermal history of the Universe. Observations made by the \COBEF satellite in the 90s confirm that the CMB spectrum is extremely close to a blackbody, with departures limited to $\Delta I/ I \lesssim 10^{-5}$ \citep{Fixsen1996, Fixsen2003}.

It is well-known that many interactions can lead to spectral distortions due to out-of-equilibrium processes \citep[e.g.,][]{Zeldovich1969, Sunyaev1970SPEC, Burigana1991, Hu1993, Chluba2011therm, Sunyaev2013, Tashiro2014, Lucca2020}. In recent years, the potential of CMB spectral distortions has led to renewed interest in this topic and experimental concepts like \PIXIE \citep{Kogut2011PIXIE, Kogut2016SPIE, KogutPIXIE2024} and beyond \citep{PRISM2013WPII, Chluba2019, Chluba2021Voyage} may open an entirely new window to the early Universe. A first detection of the largest $\Lambda$CDM distortion from the late Universe \citep{Refregier2000, Hill2015, Chluba2016} is now coming into reach with BISOU \citep{BISOU}, which has recently been moved into Phase A of its development, promising a factor of $20-30$ improvement in sensitivity over \COBEF. Additional ground-based spectrometer concepts such as APSERa \citep{Mayuri2015}, TMS \citep{Jose2020TMS} and COSMO \citep{Masi2021} will further advance the experimental frontier with different technology and in various observing bands.

The nature of the dark sector remains elusive. Amongst the more minimal theories is an extension to the standard model gauge groups by an additional $U_{\rm D}(1)$, giving us a new gauge boson $\gamma_{\rm d}$ to study \citep{Holdom1985}. The presence of this additional {\it dark photon} implies a kinetic mixing between the standard model photon, $\gamma$, and $\gamma_{\rm d}$, which can allow for energy and entropy transfer between the two sectors. This mixing can have a wide variety of phenomenological consequences (see \cite{Caputo2021} for a recent overview), even in cases where the dark photon sector is initially unpopulated.

Past works have considered the impact of kinetic mixing between a dark and standard model photon on the spectrum of the CMB, both with emphasis on oscillations between the sectors taking place after recombination \citep{Mirizzi2009a, Caputo2020} and before, during the $\mu-$ and $y-$distortion epochs \citep{Arias2012, McDermott2019}. In particular, the work performed by \citet{McDermott2019} compute the amplitude of $\mu-$ and $y-$ distortions following a Green's function approach \citep{Chluba2013Green, Chluba2015GreensII} based on the non-thermal energy transfer between the two sectors. While this is a valuable step towards understanding the induced spectral distortions, an additional subtlety not captured in their implementation of the Green's function exists when photons are directly injected or extracted from the spectrum [as opposed to a process which heats the background electrons] \citep{Chluba2015GreensII}. This subtlety (known as entropy injection/extraction) can lead to quantitatively and qualitatively different spectral features, which we demonstrate here. 

In this work, we revisit constraints placed on additional $U_{\rm D}(1)$ symmetries using improved analytic approximations, as well as the numerical thermalization code, \texttt{CosmoTherm} \citep{Chluba2011therm}. Utilization of this code ensures a robust propagation of CMB spectral distortion signatures through all relevant redshift regimes, allowing us to tease out spectral signatures that have been missed in previous setups. In addition, our treatment is no longer limited to the small distortion regime, which can become relevant for constraints derived on models leading to energy and entropy injection right after BBN \citep{Chluba2020large, Acharya2022large}.
We focus on scenarios where the dark photon sector ($\rho_{\rm d} \ll \rho_{\rm CDM}$) is initially unpopulated, and find that entropy extraction dominates over the energy extraction constraints derived in \citet{McDermott2019}, which {\it flips the sign} of the spectral distortion and tightens the constraint contours by a factor of $\simeq 1.5$. For scenarios in which $\rho_{\rm d} \simeq \rho_{\rm CDM}$, the effect of entropy injection can similarly dominate over energy injection; however, distortion constraints are usually not competitive in this regime \citep[e.g.,][]{Caputo2021}, and we defer a re-analysis to the future. We also find that a careful treatment of the large distortion evolution significantly improves the CMB distortion constraints for $10^{-4}\,\eV \lesssim \mdp \lesssim 10^{-3}\,\eV$.

The remainder of the paper is structured as follows: In Sec.~\ref{sec:level3}, we briefly discuss the physics of photon-dark photon conversions.
Following this, in Sec.~\ref{sec:level2} we review the Green's function formalism for computing spectral distortions, highlighting an approximation that can be used for entropy injection at $z\lesssim \pot{4}{6}$ (or $\mdp\lesssim 10^{-4}\,\eV$). We explain the numerical treatment of the problem using {\tt CosmoTherm} in Sect.~\ref{sec:level4b}. In Sec.~\ref{sec:level4} we illustrate the numerical solutions for the evolving distortion spectra and highlight the constraints we obtain using the CMB.
We conclude in Sec. \ref{sec:level5}. Unless stated otherwise, we use natural units where $c = \hbar = k_{\rm b} = 1$ throughout.

\vspace{-4mm}
\section{Photon-Dark Photon Conversions} \label{sec:level3}
An additional $U_{\rm d}(1)$ symmetry is a relatively minimal extension to the gauge structure of the standard model, whose gauge boson (the dark photon) is also capable of serving as a stable dark matter candidate. The possible existence of dark photons can be probed through a so-called kinetic mixing between the dark and visible sectors, whose strength is governed by the mixing parameter $\epsilon$. The typically effective Lagrangian coupling the two sectors takes the form 
\begin{align} \label{eq:eff_lagrangian}
    \mathcal{L} = -\frac{1}{4} F_{\mu \nu}F^{\mu\nu} - \frac{1}{4}B_{\mu \nu} B^{\mu\nu} + \frac{\epsilon}{2} B_{\mu\nu} F^{\mu\nu} + \frac{\mdp^2}{2} B_{\mu} B^{\mu} + j^{\mu}_{\rm em} A_{\mu}.
\end{align}
In this expression, $F_{\mu \nu}$ and $B_{\mu\nu}$ are the field strength tensors for the photon ($A_{\mu}$) and dark photon ($B_{\mu}$) respectively, $\mdp$ is the dark photon mass, and $j_{\rm em}^{\mu}$ is the electromagnetic current, which only couples to the standard model photon. We have assumed a small value of the mixing parameter $\epsilon$, and a more general forms of the Lagrangian density can be found in \citet{Jaeckel2008,Mirizzi2009a}.

It is possible to diagonalize the non-canonical kinetic term by performing a field redefinition. These transformations make it possible to define both an interaction basis, as well as a propagation basis for the two level system. The fact that these two bases are not aligned leads to the possibility of photon-dark photon mixing \citep{Okun1982}, analogous to the more familiar scenario of (vacuum) neutrino oscillations.

It is well known that the universe (especially during pre-recombination era) does not behave like a vacuum. In-medium effects modify the photon dispersion relationship, bestowing the primordial radiation with a thermal mass \citep[e.g.,][]{Jackson, Alonso2020Wondrous}, $m_{\gamma} \approx \omega_{\rm pl}$, which for temperatures below the electron-positron threshold ($T \ll 1$~MeV), is given by
\begin{align}
\omega_{\rm pl} \simeq \frac{4\pi \alpha_{\rm em} n_{\rm e}}{m_{\rm e}},
\end{align}
where $m_{\rm e}$ is the electron mass, $\alpha_{\rm em}$ is the fine-structure constant, and $n_{\rm e}$ denotes the free electron number density. Strictly speaking, the photon mass is also frequency dependent, $m^2_{\gamma} \simeq \omega^2_{\rm pl} - 2\omega^2 \left[({\rm n}-1)_{\rm H} + ({\rm n}-1)_{\rm He} \right]$, where the refractive index induced by scattering off of hydrogen and helium can serve to lower $m_{\gamma}$ at high frequencies. However, for the range of dark photon masses considered here, the negative contributions remain negligible for the bulk of the CMB, and we can simply approximate $m_{\gamma} = \omega_{\rm pl}$.

The presence of a thermal photon mass modifies the mixing angle away from its vacuum value\footnote{This is described by a thermal mixing angle, $\epsilon_{\rm T} \neq \epsilon$, whose exact details can be found in e.g. \citet{Mirizzi2009a}. For the remainder of this work, we only consider the vacuum angle ($\epsilon$), which is what appears in the Lagrangian and in the Landau-Zener expression, Eq.~\eqref{eq:adiabacity}.} \citep{Redondo2008a, Jaeckel2008}, and furthers the neutrino analogy by allowing resonant mixing (similar to the Mikheev-Smirnov-Wolfenstein effect) to occur when the matching condition $m_{\rm d} \approx m_{\gamma}(z_{\rm con})$ is satisfied. For technical details we refer the interested reader to \cite{Mirizzi2009a}. Ultimately, the probability for conversion from a photon into a dark photon (or vice-versa) near the resonance can be expressed as 
\begin{align}
    P_{\rm \gamma \rightarrow \gamma_{\rm d}} \simeq (1 - p)\,\delta(z-z_{\rm con}),
\end{align}
where the quantity $p$ is a measure of the adiabacity of the conversion process and $z_{\rm con}$ is the conversion redshift. This so-called crossing probability has been previously obtained using the Landau-Zener expression and is given by
\begin{align} \label{eq:adiabacity}
    p \simeq {\rm exp}\left( -\frac{\pi m_{\rm d}^2 \epsilon^2}{\omega \, H(z)\, (1+z)} \left| \frac{1}{\omega_{\rm pl}^2 }\frac{\id \omega^2_{\rm pl}}{\id z} \right|^{-1}\right),
\end{align}
where $\omega$ is the frequency of any given photon, and we evaluate this quantity only at the resonance point. Resonances of this type are quite narrow \citep{Mirizzi2009a}, and the direction of the conversion process will depend on the relative populations of photons and dark photons at the redshift for which the resonance is encountered. 

For this work we consider only isotropic (background) conversions, deferring a discussion of possible anisotropies to a future paper. We will also restrict ourselves to $m_{\rm d} \geq 10^{-12}$ eV to focus on highlighting primordial (i.e. pre-reionization) spectral distortions. In these limits, the photon mass decreases monotonically with redshift, following the free electron fraction. This monotonic behaviour allows us to ignore possible phase interference effects arising from multiple rapid level crossings \citep{Brahma2023} which can lower the overall conversion probability.

Assuming that initially there are {\rm no} dark photons ($\rho_{\rm d} \ll \rho_{\rm CDM}$), at $z_{\rm con}$ a fraction of the CMB photons will convert into dark photons. To understand the setup of the problem one can rewrite the conversion probability in the form
\begin{align}
\label{eq:Pcon_simp}
    P_{\rm \gamma \rightarrow \gamma_{\rm d}}(\gammacon, x) \simeq 1-\exp\left(-\frac{\gammacon}{x}\right),
\end{align}
where we used the dimensionless (and redshift independent) frequency $x=\omega/\TCMB$ to define the conversion parameter
\begin{align}
\gammacon(\epsilon, \mdp)=
\left.\frac{\pi m_{\rm d}^2 \epsilon^2}{ \TCMB(z)H(z)\, (1+z)} \left| \frac{1}{\omega_{\rm pl}^2 }\frac{\id \omega^2_{\rm pl}}{\id z} \right|^{-1}\right|_{z=\zcon}.
\end{align}
Conversion of photons into dark photons is thus very efficient at $x \ll \gammacon$. If $\gammacon \ll 1$, this means that only a small fraction of energy and entropy is removed from the photon field at the conversion redshift. However, for $\gammacon\gtrsim 1$, a significant amount of energy can be converted into dark photons (the peak of the blackbody is at $\nu \simeq 160\,\GHz$ or $x \simeq 1.2$). In this case, one also has to consider the change of the initial CMB temperature at $\zcon$, which becomes very important for setting the initial conditions especially in the large distortion regime. Indeed some or the parameter space has to be avoided since the end of the BBN / e$^+$e$^-$ era would coincide with the conversion redshift, a regime that requires a more involved treatment beyond the scope of this work. 

%
\begin{figure}
\includegraphics[width=\columnwidth]{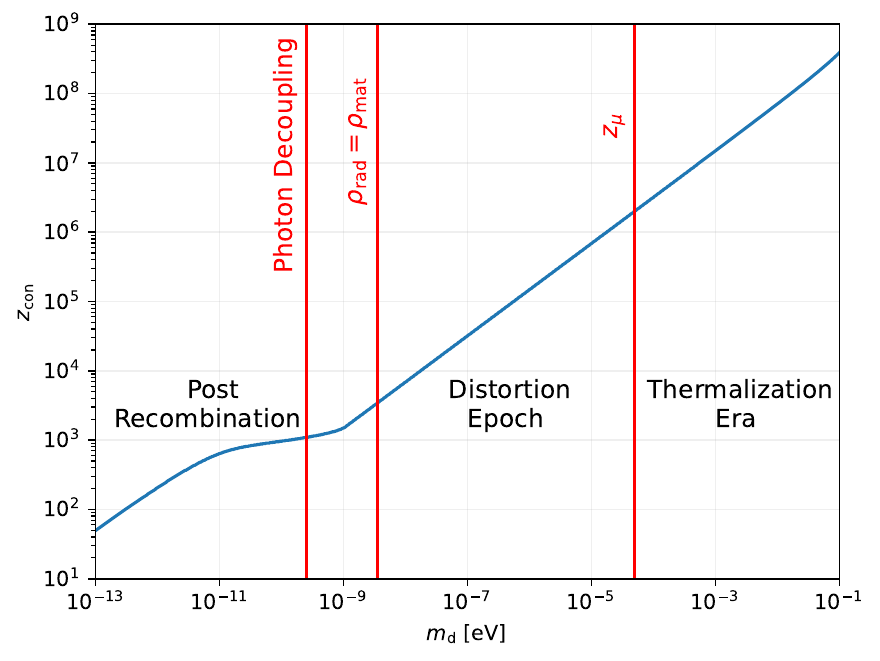}
\caption{Photon conversion redshift as a function of the dark photon mass. We also indicate some of the important mass thresholds for various eras in the history of the Universe.}
\label{fig:zcon}
\end{figure}
%
In Fig.~\ref{fig:zcon}, we show the conversion redshift as a function of the dark photon mass. This is particularly instructive as the conversion redshift is independent of $\epsilon$ and thus allows us to easily relate the dark photon mass to the type of distortion one should expect. Dark photons in the range $10^{-12} \, {\rm eV} \lesssim m_{\rm d} \lesssim 5\times10^{-10} \, {\rm eV}$ convert in the dark ages\footnote{For $10^{-14} \, {\rm eV} \lesssim m_{\rm d} \lesssim 10^{-12} \, {\rm eV}$, reionization is capable of inducing a second level crossing as the photon mass rises at $z\simeq 10$. We consider this case, as well as the effects of inhomogeneities in a future paper.} where the induced spectral distortion will remain mostly frozen in. In this case, one cannot just compare against the \COBEF residuals, but should perform a reanalysis of the data utilizing the exact spectral form of the distortion. Between approximately $5\times10^{-10} \, {\rm eV} \lesssim m_{\rm d} \lesssim 5\times10^{-5} \, {\rm eV}$, more familiar $\mu$- and $y$-type spectral distortions will be sourced as the spectrum attempts to re-thermalize through Compton interactions. However, as we show below, the dark photon distortion is always close to a $\mu$ distortion, almost independently of the conversion redshift.

At even higher masses, $m_{\rm d} \gtrsim 5\times 10^{-5}$, conversions take place in an era where all thermalization processes are rapid. Small distortions that are produced in this epoch are quickly washed away, leading to small temperature shifts. Large distortions (defined below) require a more involved treatment, and in general take longer to fully thermalize \citep{Chluba2020large}. In the following sections, we also extend spectral distortion constraints into this mass range by considering regions of parameter space where large distortions are produced. Conversions in this mass range are also subject to constraints from the effective number of relativistic degrees of freedom \citep{Jaeckel2008, McDermott2020}, and we further discuss the interplay of these constraints.

We would like to emphasise that distortions are formed regardless of whether the dark photon converts into the photon (e.g. when $\rho_{\rm d} \simeq \rho_{\rm CDM}$), or the CMB converts into the dark sector (when initially $\rho_{\rm d}\ll \rho_{\rm CDM}$). As the constraints from CMB distortions are highly competitive in the latter case, we choose to focus on the $\gamma \rightarrow \gamma_{\rm d}$ scenario in this work, leaving a discussion of more general scenarios to future analyses.

\section{Photon Injection Green's Function} \label{sec:level2}
The CMB frequency spectrum contains valuable information about the thermal history of the pre-recombination epoch that is difficult to access through other means \citep[e.g.,][]{Chluba2021Voyage}. In the very early Universe, processes such as Bremsstrahlung, Compton, and double Compton scattering\footnote{See \citet{Chluba2007a, Chluba2020BRpack, Ravenni2020DC, Sarkar2019} for precision modeling of these important processes.} are extremely efficient at maintaining a precise thermal equilibrium between the pre-recombination photons and the other components of the plasma. However, as the Universe expands these interactions begin to freeze out, after which any non-thermal energy or entropy (direct photon) injections are incapable of fully thermalizing. 

Number changing interactions (Bremsstrahlung and double Compton) become inefficient at $z\lesssim z_{\mu} \simeq 1.98 \times 10^6$, at which point the photon occupation number is driven towards a Bose-Einstein type spectrum with chemical potential $\mu$ proportional to the energy and entropy injection rates. Thus, $z_{\mu}$ marks the transition from the so-called\footnote{'$T$' for temperature, as non-thermal injections (extractions) above $z_{\mu}$ will raise (lower) the CMB monopole temperature.} $T-$ to the $\mu-$ era. At $z\lesssim \pot{5}{4}$, Comptonization also becomes inefficient and the $y$-era is entered. 

In distortion theory, one also has to distinguish the regime of small and large distortions \citep{Chluba2020large, Acharya2022large}. Due to limits from \COBEF, the evolution at $z\lesssim z_{\mu}$ is bound to be in the small distortion regime, where the problem can be fully linearized. However, at $z\gtrsim z_{\mu}$ a large energy and entropy change is in principle possible given the efficiency of turning the distorted CMB into a blackbody again. For dark photon masses $\mdp\gtrsim 10^{-4}\,\eV$, this is actually relevant and modifies the constraints as we show here. We will start our description of distortion theory for the small distortion limit and then consider the more general large distortion regime. 

\vspace{-4mm}
\subsection{Small distortion evolution}

\vspace{-1mm}
\subsubsection{$\mu$-era}
\vspace{-1mm}
In the $\mu$-era ($\pot{5}{4}\lesssim z \lesssim \pot{2}{6}$), the amplitude for an induced $\mu$ distortion can be readily approximated using the photon injection Green's function method laid out in \citet{Chluba2014, Chluba2015GreensII} [see also \citet{Bolliet2020, Cyr2023} for more recent results and applications]. Assuming a single conversion of photons into dark photon, the final $\mu$-parameter is given by
\begin{align} 
\label{eq:mu-GF}
    \mucon \simeq \frac{3}{\kappa} \,\left[\frac{\Delta \rho_\gamma}{\rho_\gamma}\Bigg|_{\rm con}  - \frac{4}{3} \frac{\Delta N_\gamma}{N_\gamma}\Bigg|_{\rm con}\right] J_{\mu}(\zcon),
\end{align}
where $\kappa \simeq 2.1419$ [i.e., $3/\kappa \approx 1.4007$] is a dimensionless numerical factor, $\epsilon_\rho=\Delta \rho_\gamma/\rho_\gamma\big|_{\rm con}$ and $\epsilon_N=\Delta N_\gamma/N_\gamma\big|_{\rm con}$ respectively denote the fractional change of the photon energy and number density with respect to the {\it initial} blackbody\footnote{These should not be confused with the coupling strength $\epsilon$, which has {\it no} subscript throughout this paper.}, and $J_{\mu}(z)$ is the $\mu$-distortion visibility function. In this work, we approximate $J_{\mu}(z)$ as \citep{Chluba2013Green}
\begin{align}
    J_{\mu}(z) \approx J_{\rm bb}(z)\left[ 1 - {\rm exp}\left( - \left[ \frac{1+z}{5.8\times 10^4}\right]^{1.88} \right) \right], \nonumber
\end{align}
where $J_{\rm bb}(z) \approx {\rm e}^{-(z/z_\mu)^{5/2}}$ is the distortion visibility function and $z_\mu$ is the redshift above which small distortions are efficiently washed out. For single conversion, this approximation should be accurate at the level of $\simeq 10\%-20\%$ during the $\mu$-era. A recent review of various approximations to the distortion visibility function can be found in \citet{Cyr2023GW}. 

The entropy term, $\epsilon_N$, in Eq.~\eqref{eq:mu-GF} is often neglected in the literature, but in principle should be included in any scenario where new photons are directly produced (or removed), such as in the case of photon-dark photon conversions. The intuition for both the sign and numerical coefficient on the entropy term are as follows. The $\mu$-era is defined as a period of the thermal history in which efficient redistribution of energy over the photon spectrum (through Compton scattering) is still active, but new photons cannot be readily created or destroyed through standard processes. When we allow photon creation or destruction through non-standard processes, relaxation to a blackbody is mediated through this additional change in $\Delta N_\gamma$, and the corresponding $\mu$ distortion is weakened. Indeed, when the fractional energy and entropy source terms during the $\mu$ era obey $\epsilon_\rho = \frac{4}{3} \epsilon_N$, one finds $\mu \approx 0$ even though thermal equilibrium was (temporarily) disrupted \citep[e.g.,][]{Chluba2015GreensII}.

Unlike the energy term in Eq.~\eqref{eq:mu-GF}, not all created photons will contribute to the entropy term. This is due to the fact that number changing interactions do still remain efficient for very low frequency photons, even throughout the $\mu$ era. As was originally explained in \cite{Chluba2015GreensII}, one can define a redshift dependent critical frequency $\omega_{\rm c}$, below which the absorption optical depth of the plasma is $\tau(\omega_{\rm c}, z)\gtrsim 1$. Photons produced below this frequency will be rapidly absorbed, heating the plasma (therefore contributing to the $\epsilon_\rho$ part), but not contributing to a net change in the entropy density. This effect is automatically taken into account by {\tt CosmoTherm} but for the analytic description we shall neglect this correction, which affects the results only marginally (weakening the amplitude of the distortion signal at the percent level) and is dominated by the uncertainties in modeling the exact critical absorption frequency.
 
The small distortion regime also implies $\gammacon\ll 1$. This means that right after the conversion we can write the distortion with respect to the {\it initial} blackbody at the temperature $\Tin>\TCMB$ as 
\begin{align} 
    \Delta n_{\rm in} \equiv n_{\rm bb}\left(\xin\right)\,\exp\left(-\frac{\gammacon}{\xin}\,\frac{\TCMB}{\Tin}\right)-n_{\rm bb}\left(\xin\right)  
    \approx - \gammacon \frac{n_{\rm bb}(\xin)}{\xin} \nonumber
\end{align}
for $\xin=\omega / \Tin> \gammacon$. Since $\Tin/\TCMB-1\ll 1$ we set $\TCMB/\Tin\approx 1$ in the conversion probability.
From this initial form of the distortion, we can compute the required energy and number density integrals in terms of numerical $G_{\rm k}$ factors, given by  $G_k=\int \frac{x^k}{{\rm e}^x-1} {\rm d} x$, meaning $G_1\approx 1.6449, G_2\approx 2.4041$ and $G_3\approx 6.4939$. One finds
\begin{subequations}
\begin{align}
\epsilon_\rho&=
\frac{\Delta \rho_\gamma}{\rho_\gamma}\Bigg|_{\rm con}
\approx-\gammacon\, \frac{G_2}{G_3}
\approx - 0.3702 \,\gammacon
\\
\epsilon_N&=
\frac{\Delta N_\gamma}{N_\gamma}\Bigg|_{\rm con}
\approx -\gammacon\, \frac{G_1}{G_2}\approx - 0.6842 \, \gammacon. 
\label{eq:frac_dN}
\end{align}
\end{subequations}
Putting things together, we then obtain
\begin{align}
\mucon\approx 0.7593\,\gammacon\,J_\mu(\zcon).
\end{align}
Here, we note that by neglecting the change in the number of photons we would naively obtain $\mucon\approx-0.5185\,\gammacon \,J_\mu(\zcon) \lesssim 0$. Therefore, the overall photon conversion distortion has the opposite sign and an amplitude that also is a factor of $\simeq 1.5$ larger. This aspect was missed in previous treatments and is also confirmed below by our full numerical calculations.

\vspace{-4mm}
\subsubsection{$y$-era}
To compute the distortion in the $y$-era ($z\lesssim \pot{5}{4}$) including entropy injection, one would have to use the more detailed photon injection Green's function with partial Comptonization \citep{Chluba2015GreensII}. In the literature, the entropy change is commonly neglected and a simple estimate for the $y$-parameter can be obtained using energy based arguments \citep{Zeldovich1969, Sunyaev1970mu}
\begin{align} 
\label{eq:y-GF}
    y \simeq \frac{\epsilon_\rho}{4} \, 
    J_{y}(\zcon)\approx -0.09255\,\gammacon\,J_{y}(\zcon).
\end{align}
The $y$-distortion visibility function is given by \citep{Chluba2013Green}
\begin{align}
    J_y \approx \left[ 1 + \left( \frac{1+z}{6\times 10^4}\right)^{2.58} \right]^{-1}. \nonumber
\end{align}
As we will see below, the approximation in Eq.~\eqref{eq:y-GF} again fails to capture the sign and amplitude of the distortion. 

It turns out that a better approximation (which is actually valid even in the $\mu$-era) can be obtained by simply ignoring the distortion shape and instead basing the estimate on the effective distortion energy density. Including the change in the photon entropy, we have the effective energy release
\begin{align}
\label{eq:limit_total}
\frac{\Delta \rho_\gamma}{\rho_\gamma}\Bigg|_{\rm dis}\approx \left[\frac{\Delta \rho_\gamma}{\rho_\gamma}\Bigg|_{\rm con}-\frac{4}{3}\,\frac{\Delta N_\gamma}{N_\gamma}\Bigg|_{\rm con}\right]\,J_{\rm bb}(\zcon).
\end{align}
Given a dark photon model, we can obtain $\Delta \rho_\gamma/\rho_\gamma\big|_{\rm dis}$ and compare this with the \COBEF (i.e., $\Delta \rho_\gamma/\rho_\gamma\big|_{\rm dis}\lesssim \pot{6}{-5}$ at 95\% c.l.) and \PIXIE limits on $\Delta \rho_\gamma/\rho_\gamma$. Indeed this approach yields consistent constraints for {\it all} dark photon masses below $\lesssim 10^{-4}\,\eV$ (see Sect.~\ref{sec:level4}).

\subsection{Large distortion regime}
We now consider the large distortion regime \citep{Chluba2020large, Acharya2022large}, which is relevant at the transition between the $T$ and $\mu$-eras. The related constraint supersedes the $N_{\rm eff}$ constraint at dark photon masses below $\mdp\lesssim \pot{2}{-4}\,\eV$ and shows clear departures from the small distortion limit (see below).

The most important aspect is that in the large distortion regime, distortions thermalize less efficiently \citep{Chluba2020large, Acharya2022large}. In these previous papers, constraints based on \COBEF and \PIXIE were considered, however, those calculations do not directly apply here since only pure heating/energy release scenarios were studied. Here, we also need to include changes to the number of photons which changes the thermalization dynamics that can be studied using {\tt CosmoTherm}. 

\subsubsection{Initial spectrum}
In the large distortion limit, one has to be more careful with the initial conditions. We start with a blackbody spectrum at a temperature $\Tin>\TCMB$. Since photons are converted into dark photons, the total energy density of the Universe in relativistic species does not change (see Appendix \ref{sec:app-A} for additional details). Consequently, the Hubble parameter or time-redshift mapping also remains unchanged.

The heat capacity of baryons and non-relativistic particles is negligible after BBN and electron-positron annihilation ($z\lesssim \pot{4}{7}\simeq \zBBN$ or a temperature of $\lesssim 10\,\keV$) even for extreme photon conversion. Assuming that the energy density of photons changes by $\epsilon_\rho=\Delta \rho_\gamma/\rho_\gamma\big|_{\rm con}$ relative to the initial blackbody at the conversion point, we then have to fulfill the condition $\Tin^4 (1+\epsilon_\rho)=\TCMB^4$.
Alternatively, we have $\Tin^4 =\TCMB^4 (1+\epsilon_{\rm CMB})$, where here we use today's CMB blackbody to determine $\epsilon_{\rm CMB}=\Delta \rho_\gamma/\rho_\gamma\big|_{\rm CMB}$. Put together this yields \citep{Chluba2020large, Acharya2022large}
\begin{align}
\frac{\Delta\Tin}{\TCMB}&=-\frac{(1+\epsilon_\rho)^{1/4}-1}{(1+\epsilon_\rho)^{1/4}},
\qquad
\epsilon_{\rm CMB}=-\frac{\epsilon_\rho}{1+\epsilon_\rho}.
\end{align}
In the small distortion limit, one readily finds $\frac{\Delta\Tin}{\TCMB}=\frac{\Tin-\TCMB}{\TCMB}\approx -\epsilon_\rho/4$ and $\epsilon_{\rm CMB}\approx-\epsilon_\rho$. However, for large conversion probability a small issue arises. At fixed $\xin=\omega/\Tin$, the photon conversion probability itself becomes temperature dependent such that $\epsilon_\rho\equiv\epsilon_\rho(\epsilon, \mdp, \Tin)$. One therefore has to solve a transcendental equation to determine $\Tin$ given the dark photon parameters $\epsilon, \mdp$. For this, we first compute the conversion redshift, $\zcon$, which is fixed by the dark photon mass and the ionization history \citep[e.g.,][]{Mirizzi2009b}. We assume that the ionization history is not affected by the conversion process. We then know $\TCMB$ at $\zcon$ and can compute
\begin{align}
\epsilon_\rho(\epsilon, \mdp, \Tin)&=-\frac{1}{G_3}\int \xin^3 \,P\left(\epsilon, \mdp, \Tin \xin  \right)\,n_{\rm bb}(\xin)\id \xin,
\end{align}
where we evaluated the conversion probability at $\omega =\Tin \xin$. 
Using Eq.~\eqref{eq:Pcon_simp} and the definition for $\gammacon$, we can rewrite the conversion probability as 
\begin{align}
P\left(\epsilon, \mdp, \Tin \xin\right)=1-\exp\left(-\frac{\gamma_{\rm con}}{\xin}\frac{\TCMB(\zcon)}{\Tin}\right)
\equiv P\left(\gammacon^*, \xin\right)
\end{align}
with $\gammacon^*=\gammacon\frac{\TCMB(\zcon)}{\Tin}$. This means that the change of the energy density is determined by a single effective parameter, $\gammacon^*$:
\begin{align}
\epsilon_\rho(\epsilon, \mdp, \Tin)\equiv \epsilon_\rho(\gammacon^*).
\end{align}
Using ${\Tin}/\TCMB(\zcon)=\gammacon/\gammacon^*$, we can therefore determine the initial temperature by solving the equation
\begin{align}
\frac{\gammacon^*}{\gammacon}&=\left[1+\epsilon_\rho(\gammacon^*)\right]^{1/4}
\end{align}
for $\gammacon^*$ (see Fig.~\ref{fig:gamma_contour}). This implies that one could in principle compute the mapping once and then simply read off the initial temperature for given $\gammacon(\epsilon, \mdp)$. However, the overall computational burden is minor and we simply solve the conditions each time.

Assuming small conversion probability (i.e., $\gamma^*_{\rm con}\ll 1$), we find\footnote{This approximation is obtained by assuming that all contributions to the integral come from $\xin>\gammacon$ such that one can use $P\left(\gammacon^*, \xin\right)\approx \frac{\gammacon}{\xin}-\frac{\gammacon^2}{2\xin^2}$.}
\begin{align}
\epsilon_\rho(\gammacon^*)&\approx
-\gammacon^*\,\frac{G_2}{G_3}\left[1-\frac{\gammacon^*}{2}\,\frac{G_1}{G_2}\right].
\end{align}
The asymptotic behaviour towards large $\gammacon^*$ is not captured by this expression. For this we find\footnote{This approximation is obtained by assuming that all contributions to the integral come from $\xin>1$ such that one can use $n_{\rm bb}\approx \expf{-\xin}$.} 
\begin{align}
\epsilon_\rho(\gammacon^*)
&\approx -1+\frac{(\gammacon^*)^2}{3} K_4(2\sqrt{\gammacon^*})
\end{align}
with $K_n(z)$ denoting the Bessel function of second kind to be very accurate. Even at $\gammacon^*\lesssim 1$, this expression is accurate to $\simeq 10\%$ precision. For $\gammacon^* \lesssim 500$, a better than $0.6\%$ match to the full numerical result is achieved using
\begin{align} \label{eq:rho_approx}
\epsilon_\rho(\gammacon^*)
&\approx \left(-1+\frac{(\gammacon^*)^2}{3} K_4(2\sqrt{\gammacon^*})\right)
\left[ 1+ \frac{0.1107}{1+0.8112\,\gammacon^*}\right].
\end{align}
This approximation was obtained by fitting the residual.

%
\begin{figure}
\includegraphics[width=\columnwidth]{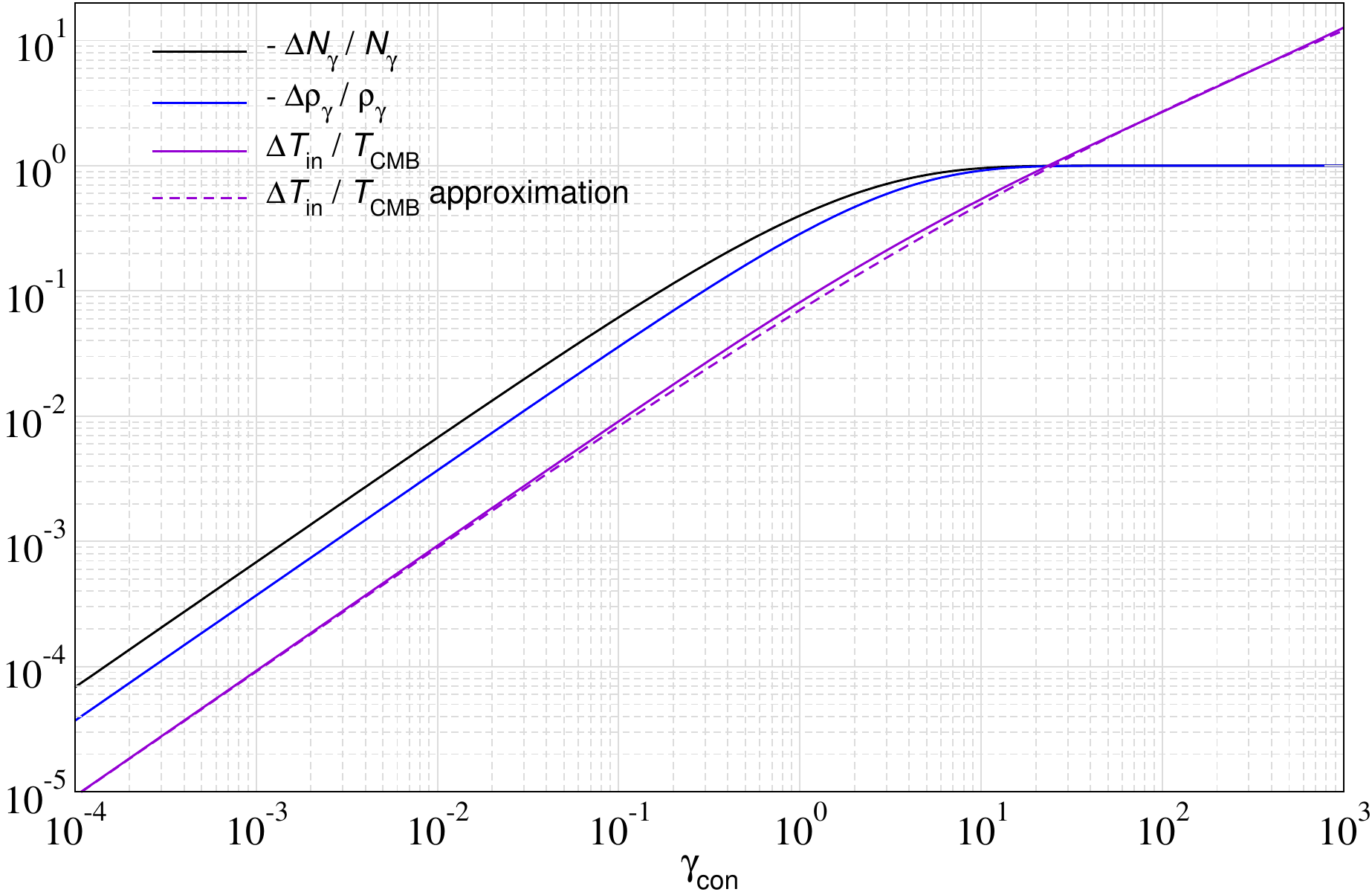}
\caption{Dependence of $\Delta \rho_\gamma/\rho_\gamma\big|_{\rm con}$, $\Delta N_\gamma/N_\gamma\big|_{\rm con}$ and $\Delta \Tin/\TCMB$ on $\gammacon$.}
\label{fig:PC_functions_gamma}
\end{figure}
%
Keeping all terms up to second order in $\epsilon_\rho$ and $\frac{\Delta\Tin}{\TCMB}$ we can also show that the initial temperature is determined by
\begin{align}
\frac{\Delta\Tin}{\TCMB}
&\approx \frac{\gamma_{\rm con}}{4}\,\frac{G_2}{G_3}\left[1+
\frac{3}{8}\left(\frac{G_2}{G_3}-\frac{4}{3}\frac{G_1}{G_2}\right)\,\gamma_{\rm con}\right]
\nonumber
\\[1mm]
&\approx 0.09255\,\gamma_{\rm con}-0.01881\,\gamma^2_{\rm con}.
\end{align}
with good convergence at $\gamma_{\rm con}\lesssim 0.2$. For $\gamma_{\rm con}\lesssim 500$, we find 
\begin{align} \label{eq:Tin_approx}
\frac{\Delta\Tin}{\TCMB}
&\approx \frac{\gamma_{\rm con}}{4}\,\frac{G_2}{G_3}\,\frac{1}{1+0.3151 \gamma_{\rm con}^{0.4426}}
\end{align}
to represent the full numerical result well, with largest departures of $\simeq 15\%$ around $\gammacon\simeq 1$. This can be useful for estimates, but in our computations we use the full numerical results from {\tt CosmoTherm}. 
%

In Fig.~\ref{fig:PC_functions_gamma} we illustrate the dependence of $\Delta \rho_\gamma/\rho_\gamma\big|_{\rm con}$, $\Delta N_\gamma/N_\gamma\big|_{\rm con}$ and $\Delta \Tin/\TCMB$ on $\gammacon$. For $\gammacon<1$, a linear scaling is found. For large $\gammacon$, both $\Delta \rho_\gamma/\rho_\gamma\big|_{\rm con}$ and $\Delta N_\gamma/N_\gamma\big|_{\rm con}$ approach $-1$, while $\Delta \Tin/\TCMB$ continues to rise as $\Delta \Tin/\TCMB \propto \gammacon^{0.56}$. In this limit most of the initial photons are found deep in the Wien tail of the initial CMB blackbody with $\Tin>\TCMB$.

%
\begin{figure}
\includegraphics[width=\columnwidth]{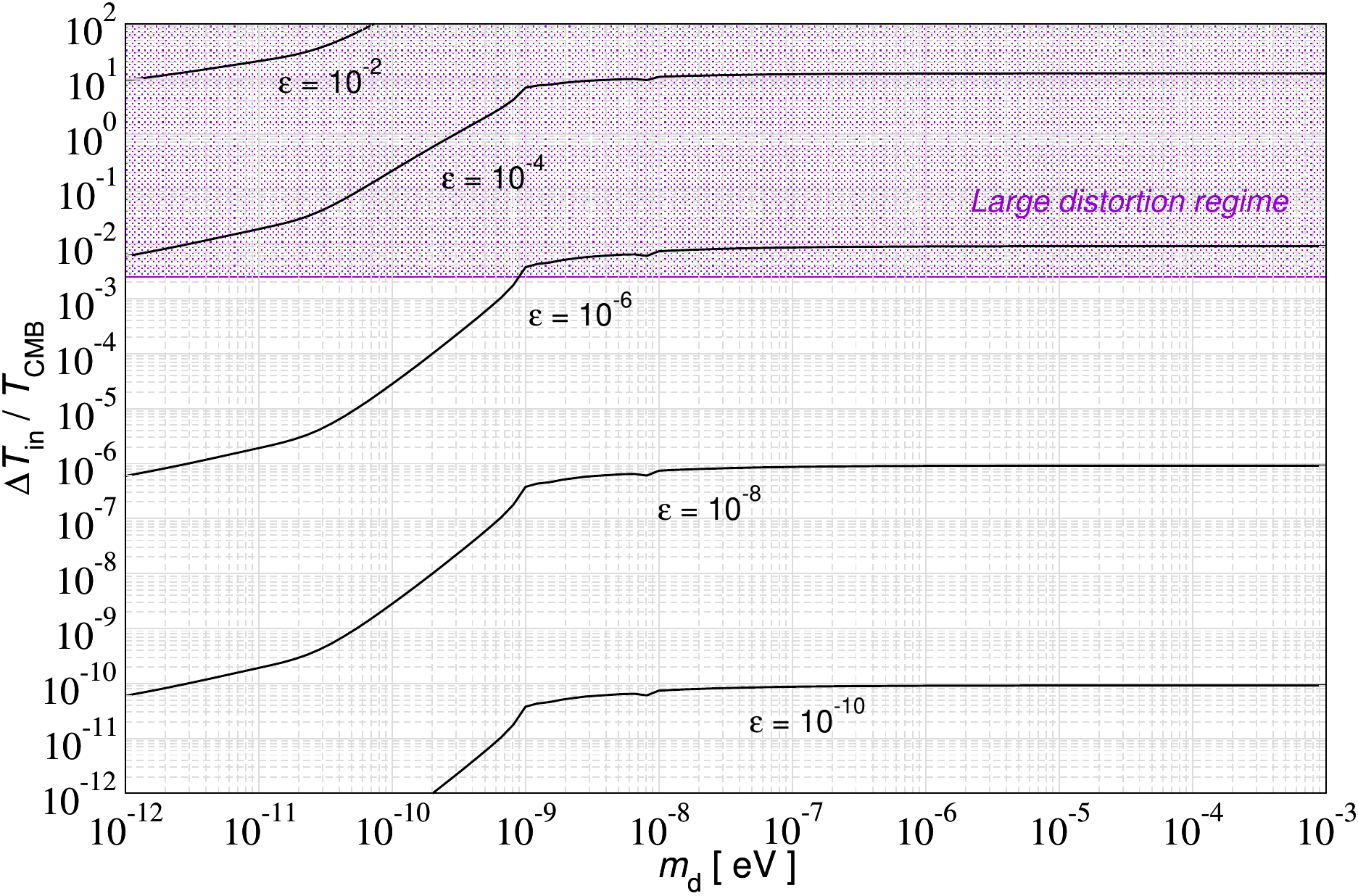}
\caption{Initial temperature change relative to the final CMB temperature for fixed values of $\epsilon$. The shaded region indicates the large distortion regime where $\epsilon_\rho>10^{-2}$.}
\label{fig:DT_TCMB_const_eps}
\end{figure}
%
In Fig.~\ref{fig:DT_TCMB_const_eps} we illustrate the dependence of the initial temperature change relative to the final CMB temperature for various values of $\mdp$ and $\epsilon$. The large distortion regime is reached for $\epsilon\gtrsim 10^{-6}$ at most masses shown in the figure. In this regime, care must be taken in setting up the initial conditions and solving the evolution equation including non-linear corrections to the Compton processes \citep{Chluba2020large, Acharya2022large}.

\subsubsection{Numerical aspects}
We can numerically solve the evolution of the spectral distortion from photon conversion using {\tt CosmoTherm} \citep{Chluba2011therm}. In the large distortion regime, a few aspects have to be kept in mind. Firstly, because one can encounter a large change in the number of CMB photons by the conversion, we use the effective temperature based on the number density of photons after the conversion as our reference temperature. Given the solution for $\frac{\Delta\Tin}{\TCMB}$, we can evaluate the number density integral to obtain $\epsilon_N
=\Delta N_\gamma/N_\gamma\big|_{\rm con}$ at the conversion redshift and relative to the initial blackbody photon number. For our computations, the precise value for $\epsilon_N$ is determined numerically using {\tt CosmoTherm}.
The initial reference blackbody is then set to a temperature
\begin{align}
\frac{T_N}{\Tin}&=(1+\epsilon_N)^{1/3}<1
\quad\text{or}\quad
\frac{T_N}{\TCMB}=\frac{(1+\epsilon_N)^{1/3}}{(1+\epsilon_\rho)^{1/4}}<1.
\end{align}
We note that $T_N<\TCMB<\Tin$. With this choice, we avoid large numerical errors due to excess photon number changes that can swamp the distortion signal. 

For late-time conversion, the constraints push the computations into the small distortion regime. However, for $\mdp\gtrsim 10^{-9}\,\eV$ and $\epsilon\gtrsim 10^{-6}$, the large distortion regime is reached (see Fig.~\ref{fig:DT_TCMB_const_eps}). The conversion redshifts for $10^{-4}\,\eV\lesssim \mdp\lesssim \pot{5}{-3}\,\eV$ fall between the beginning of the $\mu$-era and the end of BBN ($\pot{2}{6}\lesssim \zcon \lesssim \pot{4}{7}$). We define the end of BBN by assuming that the temperature has dropped below $\simeq 10\,\keV$, which ensures that electron-positron annihilation is also over. Since at these redshifts double Compton is extremely efficient, we continuously reset the reference blackbody temperature to better resolve the distortion. In addition to an energy density based criterion \citep[for details see][]{Acharya2022large} we also introduce a temperature-shift based criterion which ensures that the solution can be accurately interpolated to the new frequency grid. This is quite important at the boundaries, where too large temperature changes would lead to extrapolation errors.

To improve numerical stability, we also ensure that the evolution is initially followed very carefully by basing the step-size on the derivatives on the Comptonization $y$-parameter. Overshooting for the ODE solver is initially also suppressed to improve numerical stability and performance. We typically use $4000-8000$ grid point for the frequency discretization and include all non-linear terms in the Kompaneets equation. Relativistic corrections to Compton scattering \citep{Chluba2005Thesis, Chluba2020large} and scattering kernel corrections \citep{Acharya2021} are neglected, but the final results should be precise at the level of $1\%-10\%$ \citep[see][for details]{Chluba2020large, Acharya2022large}. We also directly compute the double Compton emissivity of the distorted photon distribution in the non-relativistic approximation \citep{Chluba2007a}.

With these settings we obtain highly accurate results in the small distortion regime. In the large distortion regime with conversion redshifts $\zcon\simeq 10^7$ or above the numerical problem becomes quite demanding. We are currently unable to explicitly compute accurate distortion signals for $\mdp\gtrsim 10^{-3}$ and $\epsilon \gtrsim \pot{2}{-5}$. In this domain, the initial temperature can exceed the present-day CMB temperature by more than one order of magnitude such that the conversion occurs at the the end of BBN and electron-position annihilation. We also expect corrections relating to the time-dependence of the conversion process to become noticeable in this domain. We therefore did not push the treatment beyond initial temperatures of $\Tin \simeq 10\,\keV$, which largely avoids these concerns. 

\section{CMB Spectral Distortion solutions} \label{sec:level4b}
We now have all the ingredients to illustrate the full evolution of the distortion for representative cases. We will start with the evolution in the large distortion regime since in this case we can visualize matters for the total photon distribution. This allows us to clarify the difference of the solution with respect to various choices of the reference blackbody spectrum. We then consider case in the $y$ and $\mu$-era and briefly discuss the late evolution in the post-recombination era.

%
\begin{figure}
\includegraphics[width=\columnwidth]{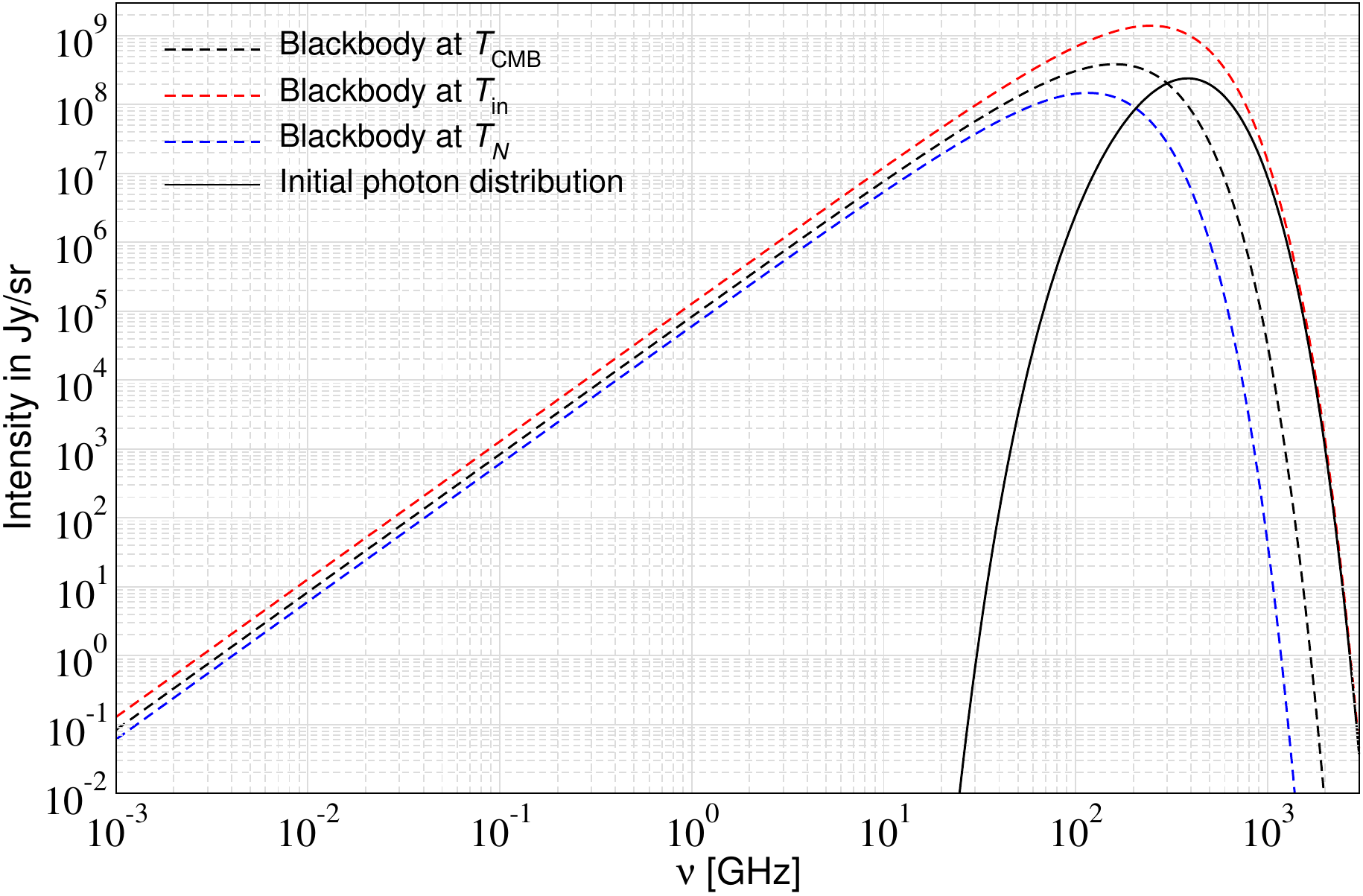}
\\[2mm]
\includegraphics[width=\columnwidth]{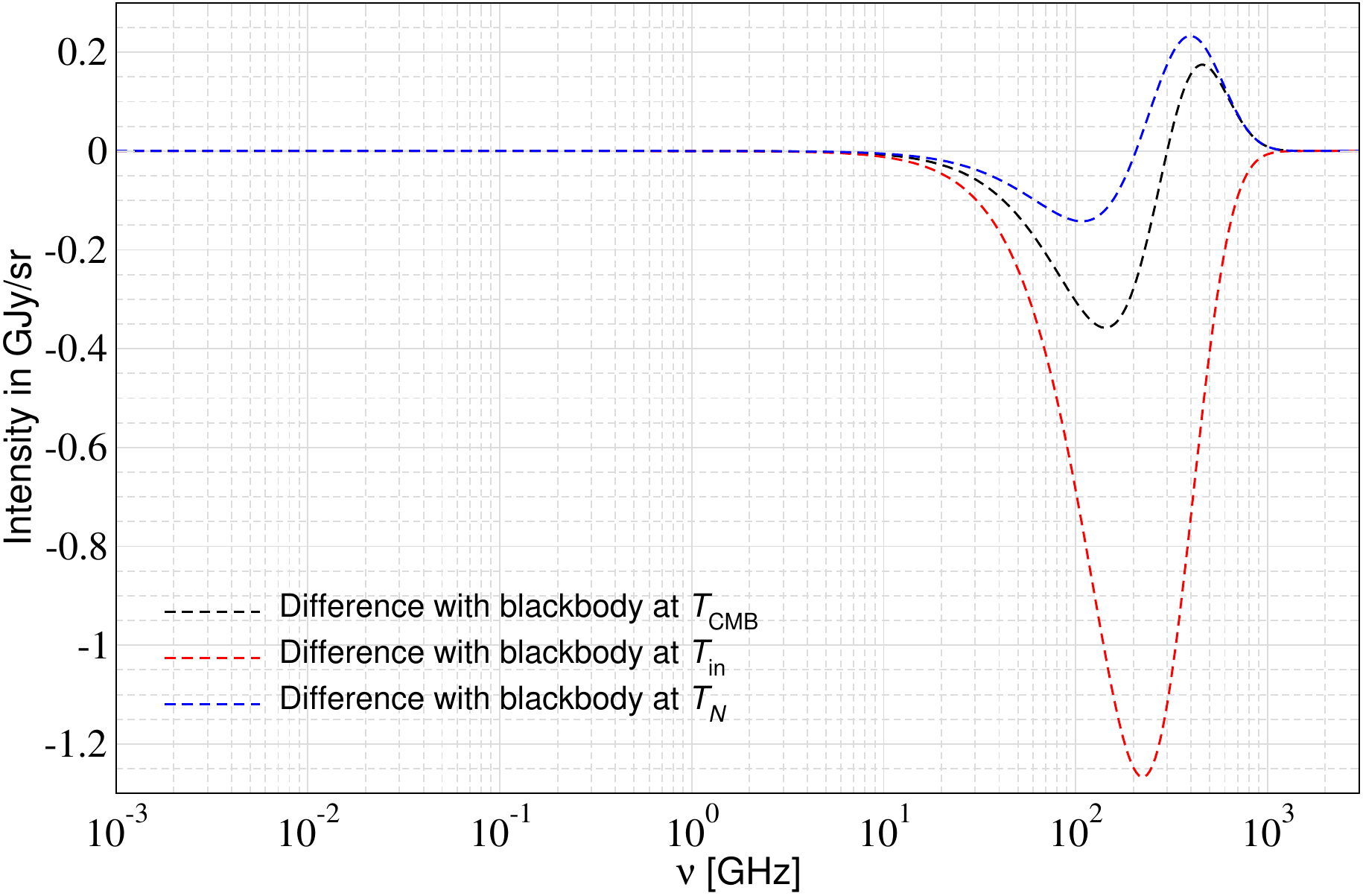}
\caption{Illustration of the initial spectrum for $\mdp=10^{-4}\,\eV$ and $\epsilon=10^{-5}$. See text for detailed discussion.}
\label{fig:Initial_large}
\end{figure}
%
\subsection{Evolution in the large distortion regime}
We shall start by considering scenarios in the large distortion regime. One case is $\mdp=10^{-4}\,\eV$ and $\epsilon=10^{-5}$, implying $\zcon\approx \pot{3.2}{6}$ and $\gammacon\approx 9.91$. This model is already ruled out by \COBEF data as it leads to $\Delta \rho_\gamma/\rho_\gamma\big|_{\rm con}\approx -0.821$ and $\Delta N_\gamma/N_\gamma\big|_{\rm con}\approx -0.894$, which does not thermalize until today. The initial temperature has $\Delta \Tin/\TCMB\approx 0.537$, implying that the initial blackbody is $\approx 1.5$ times hotter than the CMB would be today. 

In Fig.~\ref{fig:Initial_large} we illustrate the initial, distorted photon distribution in comparison to the CMB blackbody today and blackbodies at the temperatures $\Tin>\TCMB>T_N$. All spectra have been redshifted until today assuming no collision at the intermediate stages. For the considered case, the initial photon distribution is very far from that of a blackbody. The energy density is consistent with that of the CMB today, with many photons in excess of the CMB Wien tail, but the number density is significantly deficient of photons, which is also reflected in the fact that $T_N<\TCMB$. 

In the bottom panel of Fig.~\ref{fig:Initial_large}, we also illustrate the difference of the initial photon spectrum with respect to the three blackbody spectra at $\TCMB$, $\Tin$ and $T_N$. Using a reference blackbody at $T_N$ yields a distortion definition for which the number integral vanishes, while for the other choices varying amounts of photon number is carried by the distortion. For numerical applications $T_N$ is the best choice. In this case, due to non-linear terms the signal shape is similar to that of a $y$-type distortion even without Comptonization.

%
\begin{figure}
\includegraphics[width=\columnwidth]{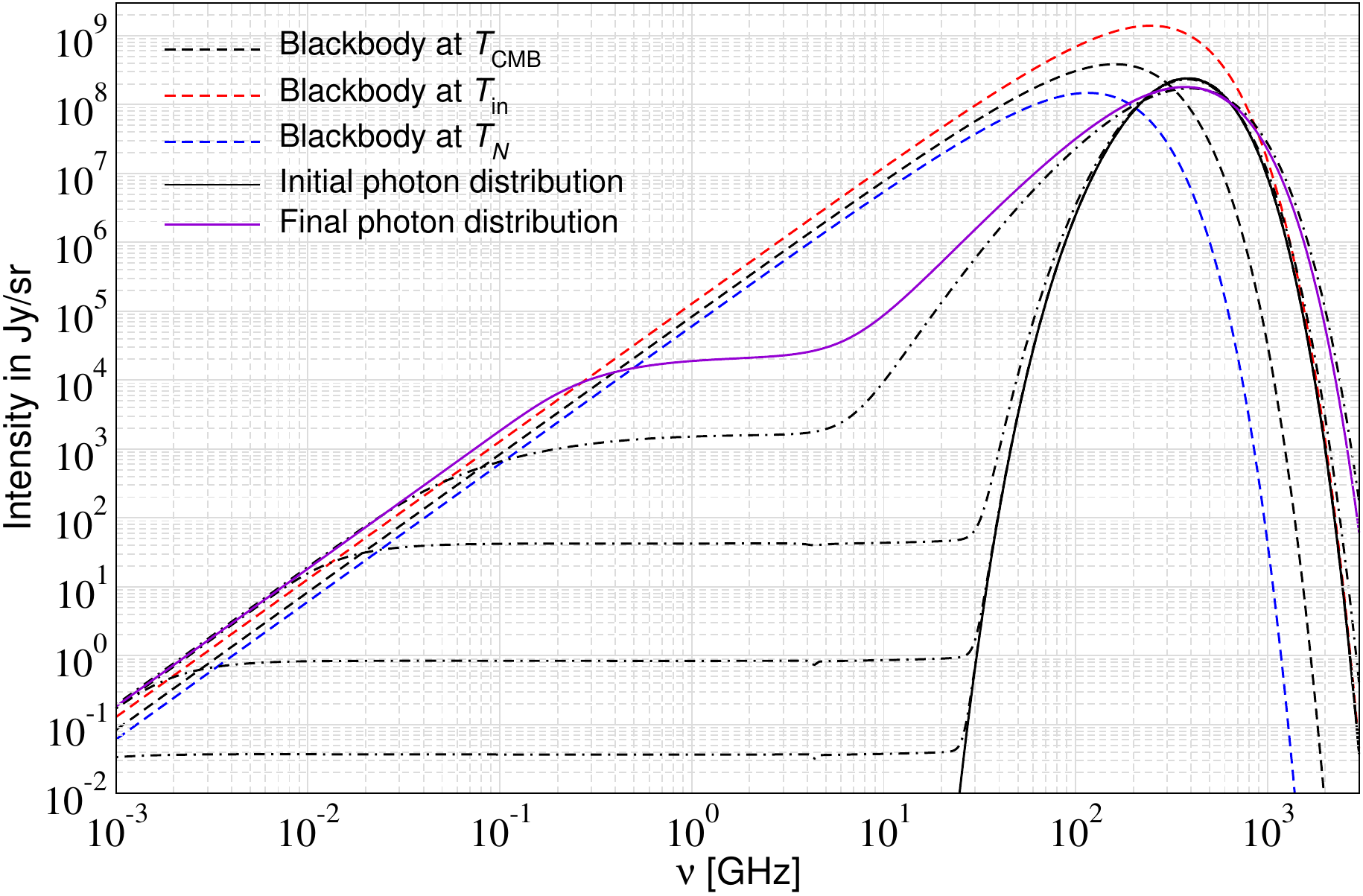}
\caption{Evolution of the spectrum for $\mdp=10^{-4}\,\eV$ and $\epsilon=10^{-5}$. The final spectrum is evaluated at $z=0$. Intermediate stages of the evolution are shown as dash-dotted lines (at arbitrary redshifts). Thermalization starts at low frequencies and gradually fills in the photon deficit of the high frequency spectrum through the combined action of double Compton, Bremsstrahlung and Compton scattering.}
\label{fig:Evolution_large}
\end{figure}
%
In Fig.~\ref{fig:Evolution_large}, we illustrate some of the intermediate evolutionary stages of the {\tt CosmoTherm} run for the large distortion scenario. The final spectrum strongly departs from that of a blackbody. 
Double Compton and Bremsstrahlung emission create a blackbody at low frequencies at the temperature of the electrons, $\Te>\Tin$, and a critical frequency that increases with time. This is because a distorted photon distribution pushes the electrons towards the Compton equilibrium temperature which for positive chemical potential exceeds that of the initial blackbody temperature \citep{Chluba2020large}. As the chemical potential diminishes, the electron temperature approaches the final CMB temperature, although in the chosen example this regime is not reached.
In the small distortion limit one would have expected a suppression of the distortion amplitude by a factor of $1/J_{\rm bb}(\zcon)\simeq 30$ due to the thermalization process. However, the evolution is inhibited in the large distortion regime.

%
\begin{figure}
\includegraphics[width=\columnwidth]{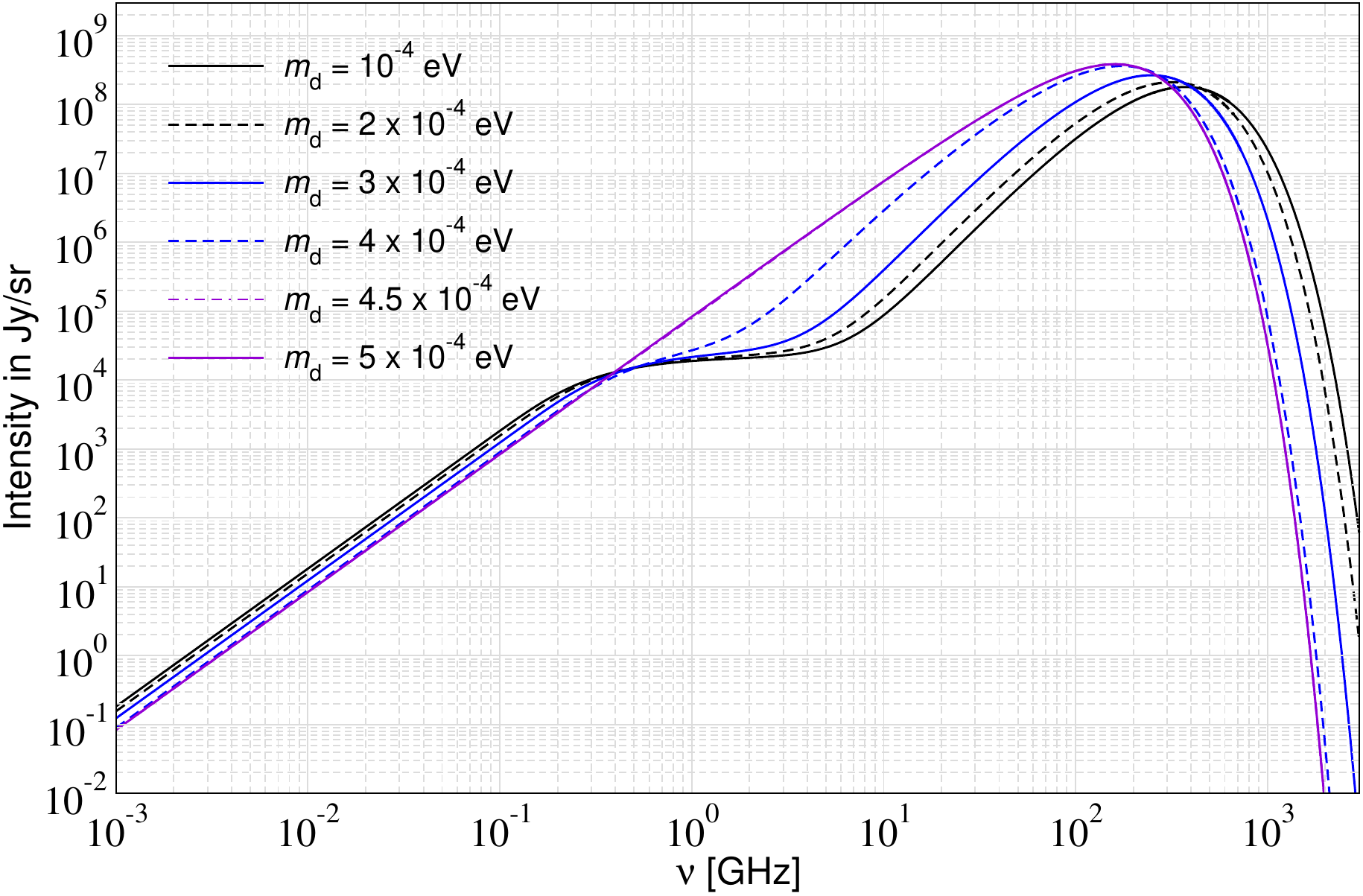}
\\[2mm]
\includegraphics[width=\columnwidth]{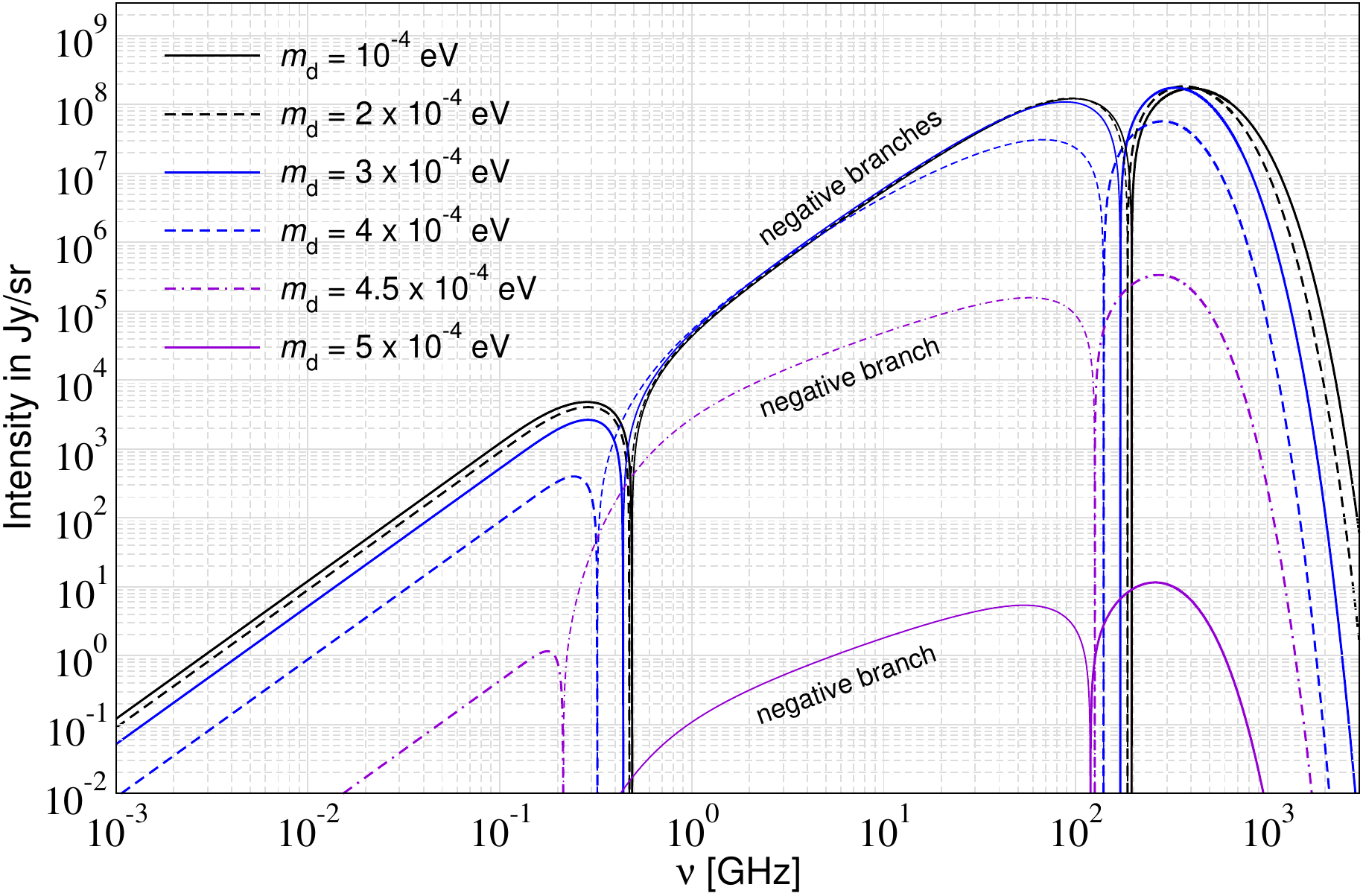}
\caption{Final spectra (upper panel) and distortions with respect to a blackbody at the momentary temperature $T_N$ (lower panel) for $\epsilon=10^{-5}$ and varying values of $\mdp$. The final spectrum is evaluated at $z=0$. For $\mdp=\pot{5}{-4}\,\eV$ this spectrum is very close to fully thermalized with a small $\mu$-type distortion.
\vspace{-3mm}
}
\label{fig:Distortion_large_var_mdp}
\end{figure}
%
We next show how the final distortion spectrum depends on $\mdp$. For $\mdp>10^{-6}\,\eV$, both $\Delta \Tin/\TCMB$ and $\Delta \rho_\gamma/\rho_\gamma\big|_{\rm con}$ are almost independent of $\mdp$. Varying $\mdp$ at constant $\epsilon$ is therefore equivalent to varying the conversion redshift only. The results for varying values of $\mdp$ are shown in Fig.~\ref{fig:Distortion_large_var_mdp}. By increasing $\mdp$ the conversion redshift increases and hence thermalization becomes more complete. 
For $\mdp=\pot{5}{-4}\,\eV$ the final spectrum is extremely close to the CMB blackbody with a remaining distortion at the level of $\Delta \rho/\rho\simeq 10^{-7}$. The conversion redshift in this case corresponds to $\zcon\approx \pot{9.3}{6}$. In the small distortion limit, one would have expected the distortion to be suppressed by a factor of\footnote{This neglects corrections to the small-distortion visibility function \citep{Khatri2012b, Chluba2014}.} $\gtrsim 10^{20}$ although here we only find a suppression by $\simeq 10^7$. This leads to an significant underestimation of the related constraint and clearly illustrates the importance of treating the large distortion regime carefully.

\subsection{Evolution across the $\mu$ and $y$-eras ($10^3\lesssim z \lesssim \pot{2}{6}$)}
The following picture for the evolution of the distortion during the $\mu$ and $y$-eras is expected: at $z\gtrsim 10^4$ a noticeable redistribution of photons by Compton scattering occurs, leading to a gradual transformation of the photon distribution from the initial state to a $\mu$-type distortion. The higher the mass of the dark photon the more complete will this transformation be. At $z\lesssim 10^4$, one can expect a small amount of Comptonization and hence the initial distortion  mainly evolves through Bremsstrahlung emission at low frequencies, leading to a small cooling of the plasma. The transition between these two regimes happens at $\mdp\simeq \pot{2}{-8}\,\eV$.

At $z\gtrsim \pot{2}{5}$ one furthermore expects low frequency photons created by the Bremsstrahlung and double Compton processes to be up-scattered towards high frequencies and help diminish the amplitude of the $\mu$-distortion \citep{Chluba2015GreensII}. Efficient thermalization is then reached at $z\gtrsim z_\mu$, where most of the distortion is converted into a change of the temperature with $T_N\rightarrow \TCMB$.

\begin{figure}
\includegraphics[width=\columnwidth]{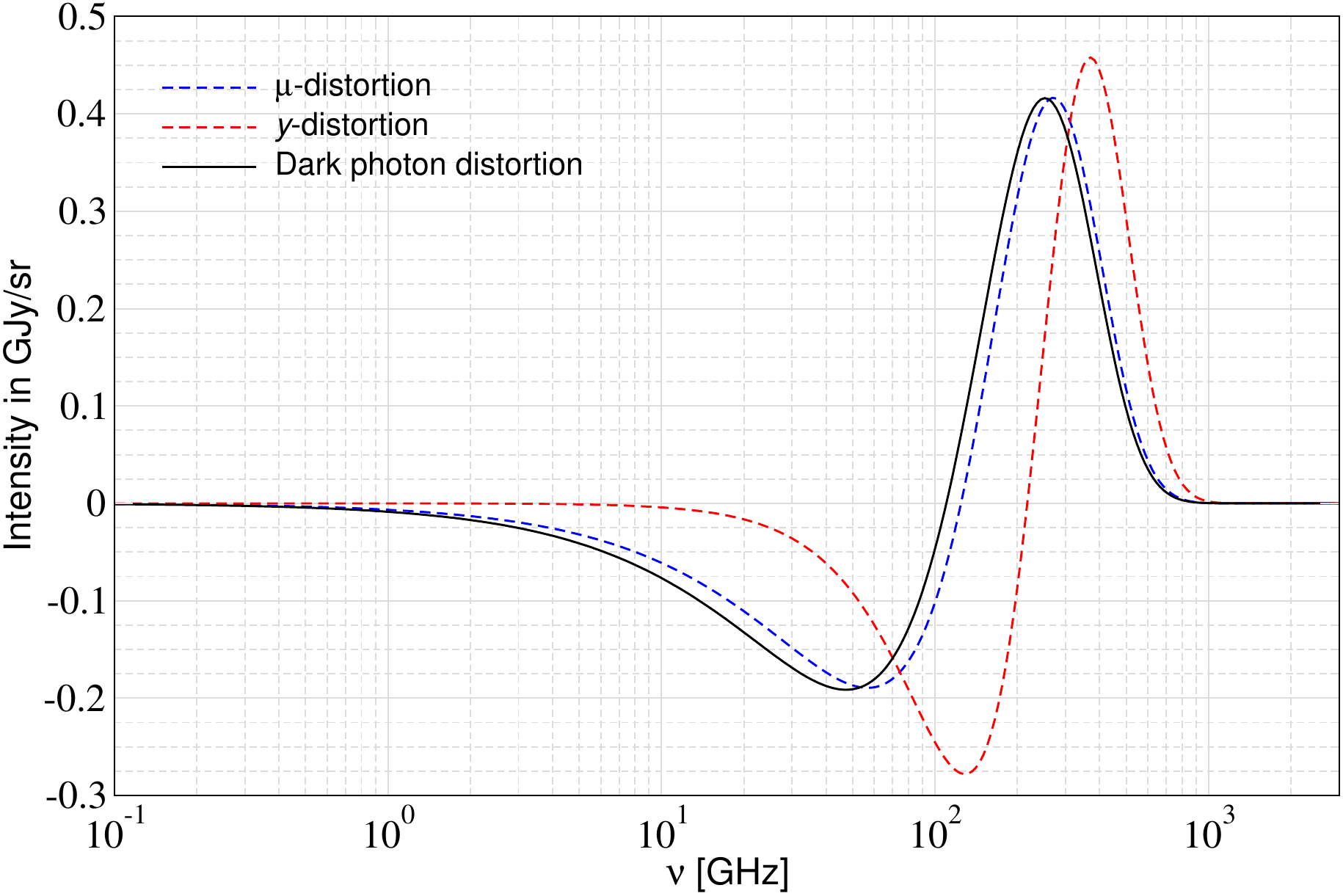}
\caption{Normalized distortion spectra $M^*(x)=1.401\,M(x)$, $Y^*(x)=Y(x)/4$ and $D^*(x)=1.845\,D(x)$ [with $1.845=1/0.5421$]. The dark photon distortion is very close to that of a $\mu$-distortion although with slightly lower crossover frequency.}
\label{fig:D_distortion}
\end{figure}

To illustrate the key effects, let us first discuss the initial spectra. We shall assume the small distortion limit, which essentially means we have to restrict ourselves to $\epsilon \lesssim \pot{5}{-7}$ almost independent of mass for the relevant redshifts. This also implies $\gammacon\lesssim 0.03$, which means that significant conversion occurs well into the Rayleigh-Jeans tail of the CMB. In this case, we can simply approximate the initial spectrum at first order in small quantities as
\begin{align}
\label{eq:n_dp}
n(\xin)&\equiv n_{\rm bb}(\xin)\exp\left(-\frac{\gammacon}{\xin}\frac{\TCMB}{\Tin}\right)
\nonumber \\
&\approx n_{\rm bb}(x)+G(x)\,\frac{\Delta \Tin}{\TCMB}-\frac{n_{\rm bb}(x)}{x}\,\gammacon
\end{align}
at $x\gtrsim \gammacon$. For convenience, we introduce the standard spectra 
\begin{subequations}
\begin{align}
G(x)&=\frac{x \expf{x}}{(\expf{x}-1)^2},
\\
Y(x)&=G(x)\big[x\coth(x/2)-4\big],
\\
M(x)&=G(x)\left[\frac{1}{\beta_M}-\frac{1}{x}\right],
\end{align}
\end{subequations}
with $\beta_M=2.1923$. The combination of the last two terms in Eq.~\eqref{eq:n_dp} are indeed close to a $\mu$-type distortion itself, as we will illustrate below (see discussion around Fig.~\ref{fig:D_distortion}). 
Using $\Delta \Tin/\TCMB\approx G_2 \gammacon/[4 G_3]$, we can write the distortion with respect to the CMB blackbody as
\begin{align}
\label{eq:Dn_dp}
\Delta n(x)&\equiv
n(\xin)-n_{\rm bb}(x)
\nonumber \\
&\approx \left[\frac{G_2}{4\,G_3}\,G(x)-\frac{n_{\rm bb}(x)}{x}\right]\,\gammacon
\nonumber \\
&\equiv \left\{\left[\frac{G_2}{4\,G_3}-\frac{G_1}{3\,G_2}\right]\,G(x)
+\left[\frac{G_1}{3\,G_2}\,G(x)-\frac{n_{\rm bb}(x)}{x}\right]\right\}\gammacon.
\end{align}
In the last line, we separated the photon number carrying term to define the dark photon distortion spectrum
\begin{align}
\label{eq:Dn_dp_N}
D(x)
&=\left[\frac{G_1}{3\,G_2}\,G(x)-\frac{n_{\rm bb}(x)}{x}\right]
\end{align}
in a photon number-conserving way, i.e., $\int x^2 D(x)\id x=0$. The number carrying (temperature shift) term $\propto G(x)$ in the last line of Eq.~\eqref{eq:Dn_dp} is not directly constrainable using the average distortion measurements as it can be absorbed in the uncertaintly of the CMB monopole temperature. However, this contribution may become constrainable using anisotropic distortion treatments \citep{Chluba2023I, Chluba2023II, Kite2023III} as a direct correlation between a $\mu$-type distortion and temperature perturbations is expected.

In the small distortion limit, $D(x)$ is the typical spectral distortion at high frequencies and almost independent\footnote{For $\gammacon \gtrsim 0.01$ some corrections from higher order frequency terms are noticeable but we confirmed that at $\epsilon \lesssim 10^{-7}$ these become subdominant.} of the value of $\gammacon \lesssim 0.03$. For dark photon models with the same value for $\gammacon$ differences in the final distortion shape and constraint can thus only arise from the variation of the dark photon mass, which directly controls the conversion redshift.
%
%
For the energy integral, we find $$\int x^3 D(x)\id x=\frac{4 G_1 G_3}{3 G_2}-G_2\approx 0.5421\,G_3.$$
For comparison, one has the integrals $\int x^3 M(x)\id x=G_3/1.401$ and $\int x^3 Y(x)\id x=4 G_3$ for a $\mu$ and $y$-distortion respectively. In Fig.~\ref{fig:D_distortion} we show the (energy) normalized distortion spectra such that $\int x^3 D^*(x)\id x=G_3$, $\int x^3 M^*(x)\id x=G_3$ and $\int x^3 Y^*(x)\id x=G_3$. The dark photon distortion is initially very close to the spectrum of a $\mu$ distortion even without any scattering effects. By equating the related energy densities we have
\begin{align}
\label{eq:mu_eff}
\gammacon\,\int x^3 D(x)\id x=\mu^*\int x^3 M(x)\id x
\end{align}
which implies $\mu^*= 0.7593\,\gammacon$. This effective $\mu$-parameter is in agreement with the estimates from Sect.~\ref{sec:level2}. It also indicates that the simple distortion energy density based constraints, Eq.~\eqref{eq:limit_total}, can be expected to work extremely well.

Given how close the dark photon distortion is to that of a $\mu$-type distortion, $D^*(x)\approx M^*(x)$, one can expect very little evolution of the distortion shape at $z\lesssim \pot{2}{6}$. This is because $M$ is a stationary solution of the Kompaneets operator \citep{Chluba2023I}, meaning the distortion shape is frozen unless extra energy is added. At high frequencies, this is indeed true and the only visible evolution occurs at low frequencies where Bremsstrahlung is still efficient. We explicitly confirmed this picture using {\tt CosmoTherm}.

\subsection{Late evolution ($z \lesssim 10^3$)}
The late evolution is even more simple, since Comptonization can be largely neglected and only the free-free process has to be treated \citep{Chluba2014}. This means that any photon conversion in this regime essentially leads to a CMB distortion given by
\begin{align}
\label{eq:Dn_dp_late}
\Delta n(x)
&\approx -0.1355\,\gammacon\,G(x)+\gammacon\, D(x)
\nonumber\\
&\approx -0.5421\,\gammacon\,G^*(x)+0.5421\,\gammacon\, D^*(x),
\end{align}
at frequencies of interest for CMB spectrometers. In the second line we expressed the distortion with respect to the normalized spectra ($\int x^3 D^*(x)\id x=\int x^3 G^*(x)\id x=G_3$) to highlight that both terms contribute equal but opposing amounts to the energy density of the distortion. One therefore has $\int x^3 \Delta n \id x \approx 0$ as required. 
At low frequencies, additional evolution occurs which may also affect 21cm fluctuations \citep{Acharya2023SPH, Cyr2024SPH}; however, the CMB constraints strongly limit these effects. A more detailed discussion is beyond the scope of this paper and overall we can expect the distortion shape to basically be given by Eq.~\eqref{eq:Dn_dp_late} at $z\lesssim 10^6$.

%
\begin{figure}
\includegraphics[width=\columnwidth]{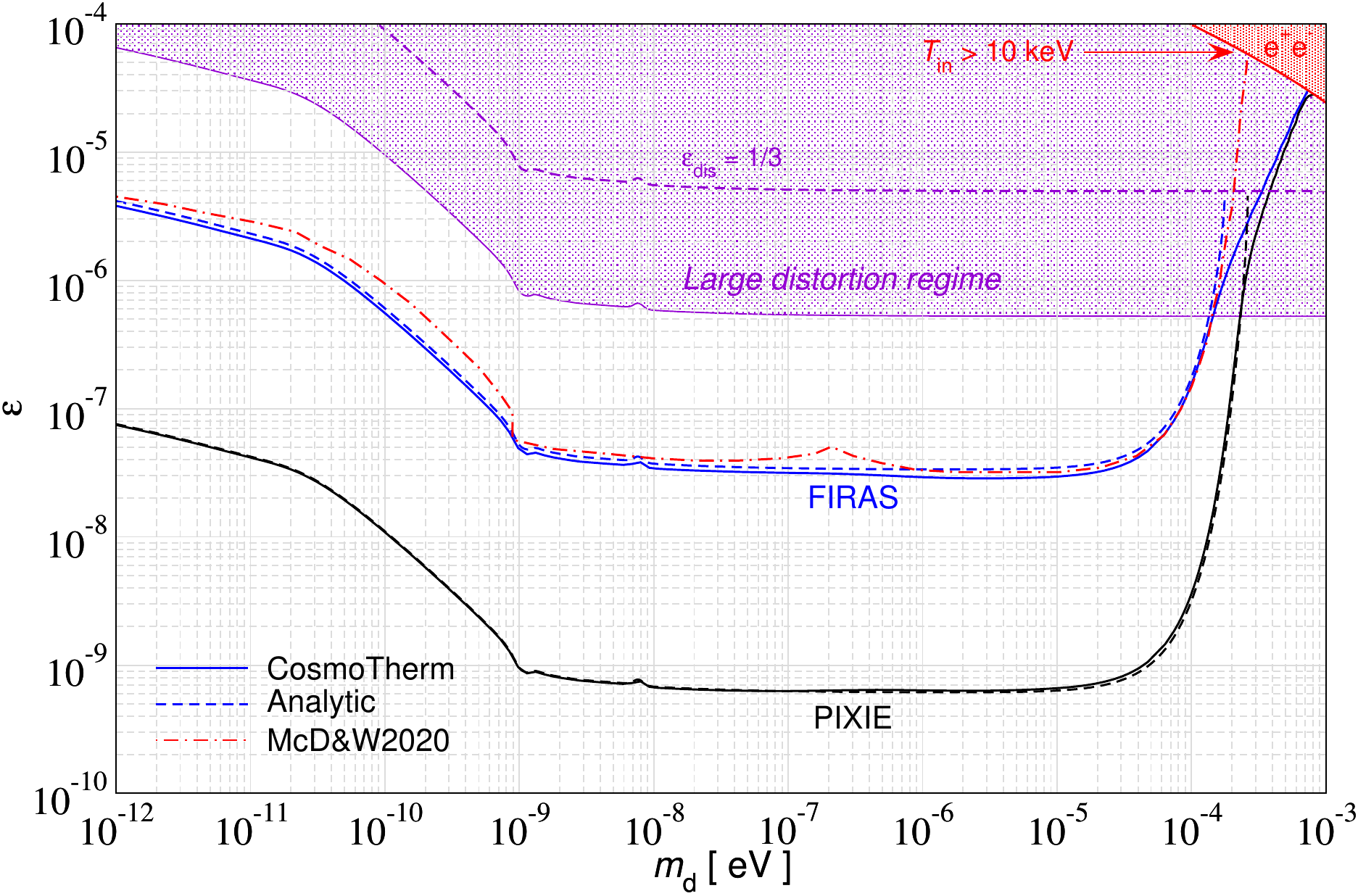}
\caption{CMB spectral distortion limits (95\% c.l.) from \COBEF and \PIXIE. The solid lines show those limits obtained directly with {\tt CosmoTherm} while the dashed lines show the estimates based on Eq.~\eqref{eq:limit_total} in the small distortion limit. The \COBEF limits given by \citet{McDermott2020} (red line) are shown for comparison. In the upper right corner we also highlight the region of parameter space for which the initial temperature at the conversion redshift exceeds $10\,\keV$ such that a clear separation with BBN / e$^+$e$^-$ annihilation is not guaranteed.}
\label{fig:Limits_distortion}
\end{figure}
%
\section{CMB Spectral Distortion Constraints on Dark Photons} \label{sec:level4}
We are now in the position to obtain constraints from \COBEF \citep{Fixsen1996} and a version of \PIXIE \citep{Kogut2016SPIE}. To obtain the precise {\tt CosmoTherm} constraints we ran 20,000 models on a log-log grid in $\mdp\in [10^{-12}, 10^{-3}]\,\eV$ and $\epsilon\in [10^{-10}, \pot{3}{-5}]$. This covers scenarios with conversion right after the end of BBN and just before the reionization process starts. We then used the obtained spectral distortions, $\Delta I_\nu$, to constrain the models. 

For \COBEF, we used the dataset provided in \citet{Fixsen1996}. We deproject both a temperature shift and spectral template for the galactic dust emission to mimic the marginalization over both the temperature and galactic emission uncertainty. For the deprojection we simply use discrete intensity vectors, $\Delta I_i=\Delta I_{\nu_i}$, $G_i=G_{\nu_i}$ and $g_i=g_{\nu_i}$ for the distortion, temperature shift and galactic emission spectra evaluated at the central frequencies, $\nu_i$, of each band. We define the scalar product $\vek{a}\cdot\vek{b}=\sum_i  a_i \, C^{-1}_{ij}\,b_j$, with $C^{-1}_{ij}$ is the inverse noise convariance matrix. We assume diagonal noise covariance using the noise levels given in \citep{Fixsen1996}. This allows us to the define the residual signal vector 
$\Delta S_i=\Delta I_i-\Delta_T G_i-\Delta_g\,g_i$ by solving the system 
\begin{subequations}
\begin{align}
\vek{\Delta I}\cdot \vek{G}&=\Delta_T \,\vek{G}\cdot \vek{G}+\Delta_g \,\vek{g}\cdot \vek{G}
\\
\vek{\Delta I}\cdot \vek{g}&=\Delta_T \,\vek{G}\cdot \vek{g}+\Delta_g\, \vek{g}\cdot \vek{g}
\end{align}
\end{subequations}
for the coefficients $\Delta_T$ and $\Delta_g$. The residual distortion signal has the property $\vek{\Delta S}\cdot \vek{G}=\vek{\Delta S}\cdot \vek{g}=0$. We then compare the obtained $\Delta S_i$ with distortion data from \citep{Fixsen1996} using a simple Gaussian likelihood to obtain the constraints on the models. 

For \PIXIE, we follow a very similar approach. We assume an effective error level $\sigma_i = 5\,{\rm Jy \,sr^{-1}}$ with $\nu\in [30, 1000]\,\GHz$ distributed in linear frequency bins of $\Delta \nu=15\,\GHz$. This is a typical setup that allows one to obtain an average $\mu$-distortion with amplitude $\mu \simeq \pot{1.4}{-8}$ at $1\sigma$ sensitivity \citep{Chluba2013PCA}, assuming that penalties from foreground marginalization are worked into the effective noise level. More detailed treatments of the foreground problem are beyond the scope of this work but have been considered elsewhere \citep{abitbol_pixie, Rotti2021}. For \PIXIE, we deproject a temperature shift and also a $y$-type distortion, since at the sensitivity of \PIXIE the large contribution from low redshifts to the average $y$-parameter \citep{Hill2015} eliminates the possibility to place tight constraints using this part of the signal \citep{Chluba2013PCA, Chluba2013fore, Cyr2023GW}.

In Fig.~\ref{fig:Limits_distortion} we give the 95\% c.l. distortion limits on the dark photon parameter space from \COBEF and \PIXIE. We show the limits obtained directly by solving the full distortion evolution using {\tt CosmoTherm} and compare to those obtained using the simple analytic estimate from Eq.~\eqref{eq:limit_total}. 
For \COBEF we assumed $|\Delta\rho/\rho|<\pot{6}{-5}$ (95\% c.l.) and for the \PIXIE setup we used $|\Delta\rho/\rho|<\pot{2}{-8}$ (95\% c.l.) when using Eq.~\eqref{eq:limit_total}. 
Aside from the constraints at masses $\mdp > 10^{-4}\,\eV$ the simple analytic approach agrees with the detailed computation at the level of $\simeq 10\%$ for \COBEF and better for \PIXIE. For \COBEF, the late-time mismatch is explained by the fact that the distortion data, taken as it is, does not account for additional systematic effects. This effectively yields a slightly tighter limit on $|\Delta\rho/\rho|$, corresponding to $|\Delta\rho/\rho|<\pot{5.3}{-5}$ (95\% c.l.) for a $\mu$-distortion \citep[see Appendix A of][]{Bolliet2020PI}, which is close to the photon conversion distortion shape.
%

At $\mdp > 10^{-4}\,\eV$, large distortion corrections become important and indeed the analytic approximation cannot be used above the line labeled with $\epsilon_{\rm dis}=1/3$, since it naively requires all the photons of the initial blackbody at $\Tin$ to convert. The treatment of large distortions allows us to extend the limit to $\mdp \simeq 10^{-3}\,\eV$. Above this mass, the distortion constraints run into the domain that would imply the conversion to occur within the BBN / e$^+$e$^-$ annihilation era, since the initial CMB temperature exceeds $\simeq 10\,\keV$ (see upper right corner in Fig.~\ref{fig:Limits_distortion}). We excluded this regime from our considerations, as a much more detailed treatment would be required.

For comparison, we also show the distortion limits given by \citet{McDermott2020}. These limits were derived omitting the effect of entropy extraction and thus produce an opposite sign of the distortion while also underestimating the limit. The limits in the large distortion regime are similarly underestimated. Lack of detail about the likelihood evaluation make a further direct comparison with our results more difficult. We therefore do not go further.

%
\begin{figure}
\includegraphics[width=\columnwidth]{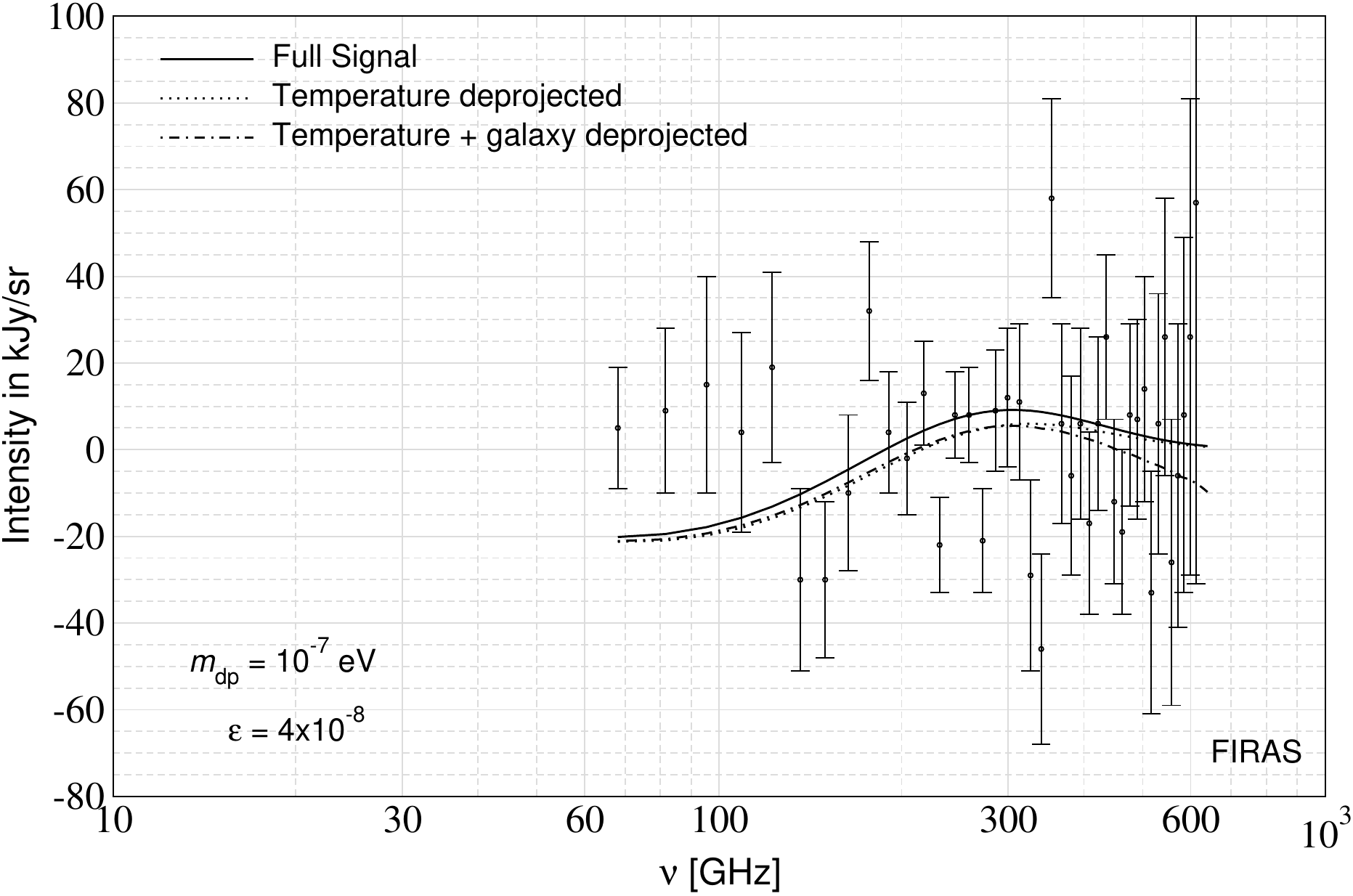}
\\[2mm]
\includegraphics[width=\columnwidth]{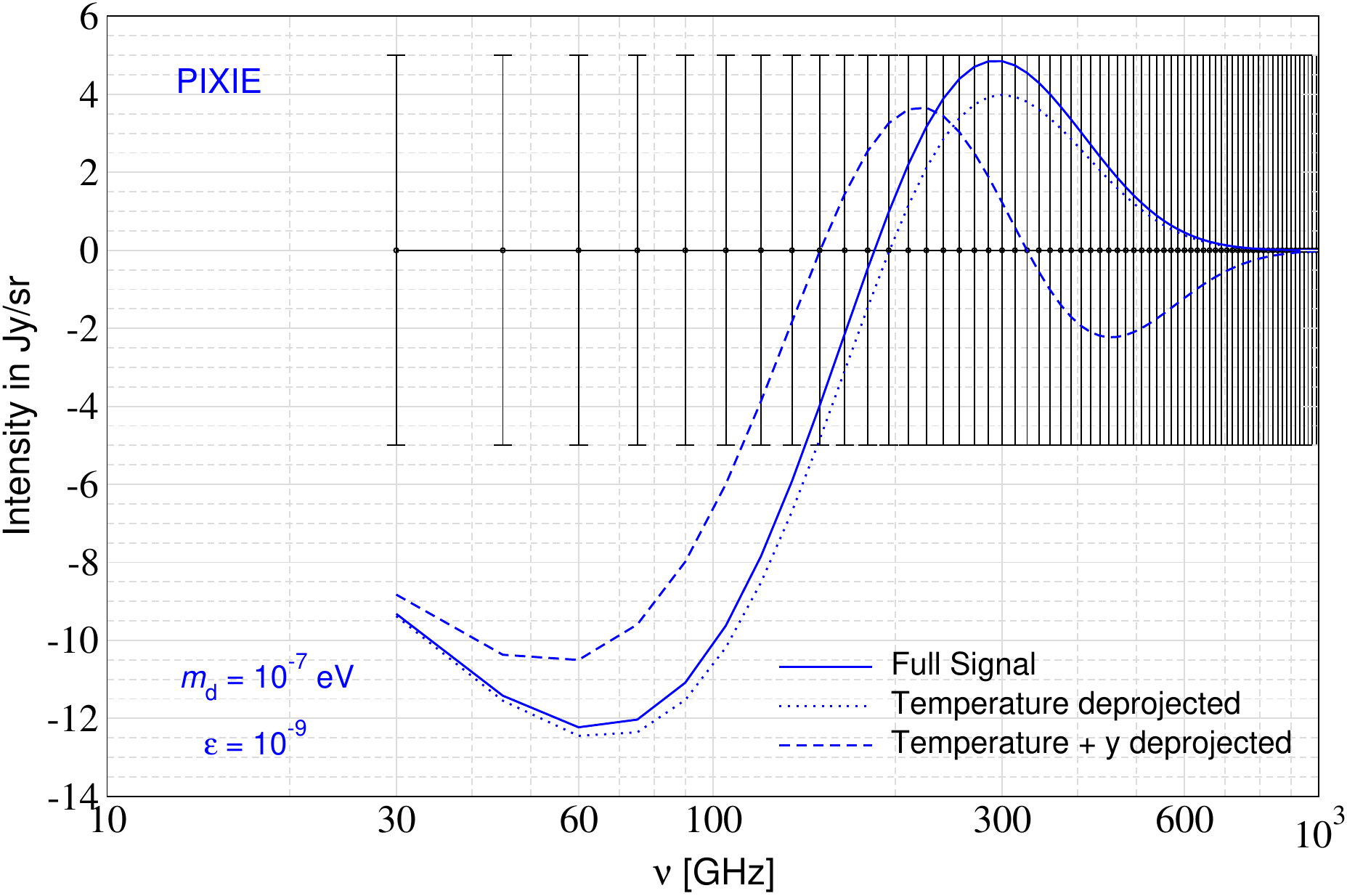}
\caption{CMB spectral distortion signals with various deprojections for \COBEF (top panel) and \PIXIE (lower panel). For illustration we also show the error and their central values. The chosen scenarios are both excluded at a few standard deviations by the respective experiment.}
\label{fig:deproj}
\end{figure}
%

\subsection{Effects of marginalization} 
\label{sec:marginalization}
In Fig.~\ref{fig:deproj}, we compare the distortion signals after various steps of deprojection for \COBEF and the \PIXIE-like setup. These deprojections mimic the effects of parameter marginalization. We can clearly see that with progressive deprojections the remaining distortion signal reduces in amplitude, thereby slightly diminishing the constraining power of the signal. The temperature deprojection is inevitable for any distortion analysis. The galactic deprojection for \COBEF modifies the remaining signal mostly in the high frequency channels where the noise level is also higher. Generally we found that the galactic deprojection did not change the final constraints by more than a few percent.

For \PIXIE, similar statements apply. However, the deprojection of $y$ does reduce the average amplitude of the remaining distortion noticeably (see lower panel of Fig.~\ref{fig:deproj}) and the final constraint is indeed weakened by\footnote{Since $\Delta \rho_\gamma/\rho_\gamma \propto \gammacon\propto \epsilon^2$ the reduction of the distortion amplitude variation propagates with a factor of 1/2 to the constraint on $\epsilon$.} $\simeq 12\%$. By combining with Sunyaev-Zeldovich cluster and X-ray measurements it may be possible to eliminate part of the low-redshift average $y$-contribution to mitigate the related loss \citep[see discussion in][]{Cyr2023GW}; however, a more detailed foreground marginalization will be required to estimate the expected improvements. 
We mention that at the level of $\Delta \rho/\rho \simeq 10^{-8}$ the dissipation of small-scale density perturbations is also expected to contribute \citep{Chluba2012, Chluba2012inflaton, Cabass2016, Chluba2016} such that in principle a joint marginalization over power spectrum parameters is required. In the context of $\Lambda$CDM, we do not expect this to pose a serious issue, while in the presence of enhanced small scale power \citep{Chluba2012inflaton} this will be relevant.
Finally, to push the limits significantly below $\Delta \rho/\rho \simeq 10^{-8}$ will also require a modeling of the expected adiabatic cooling distortion \citep{Chluba2005Thesis, Chluba2011therm, Khatri2011BE}, which can be predicted accurately and thus should not hamper the constraints much.
All these issues are left to future work.

%
\begin{figure}
\includegraphics[width=\columnwidth]{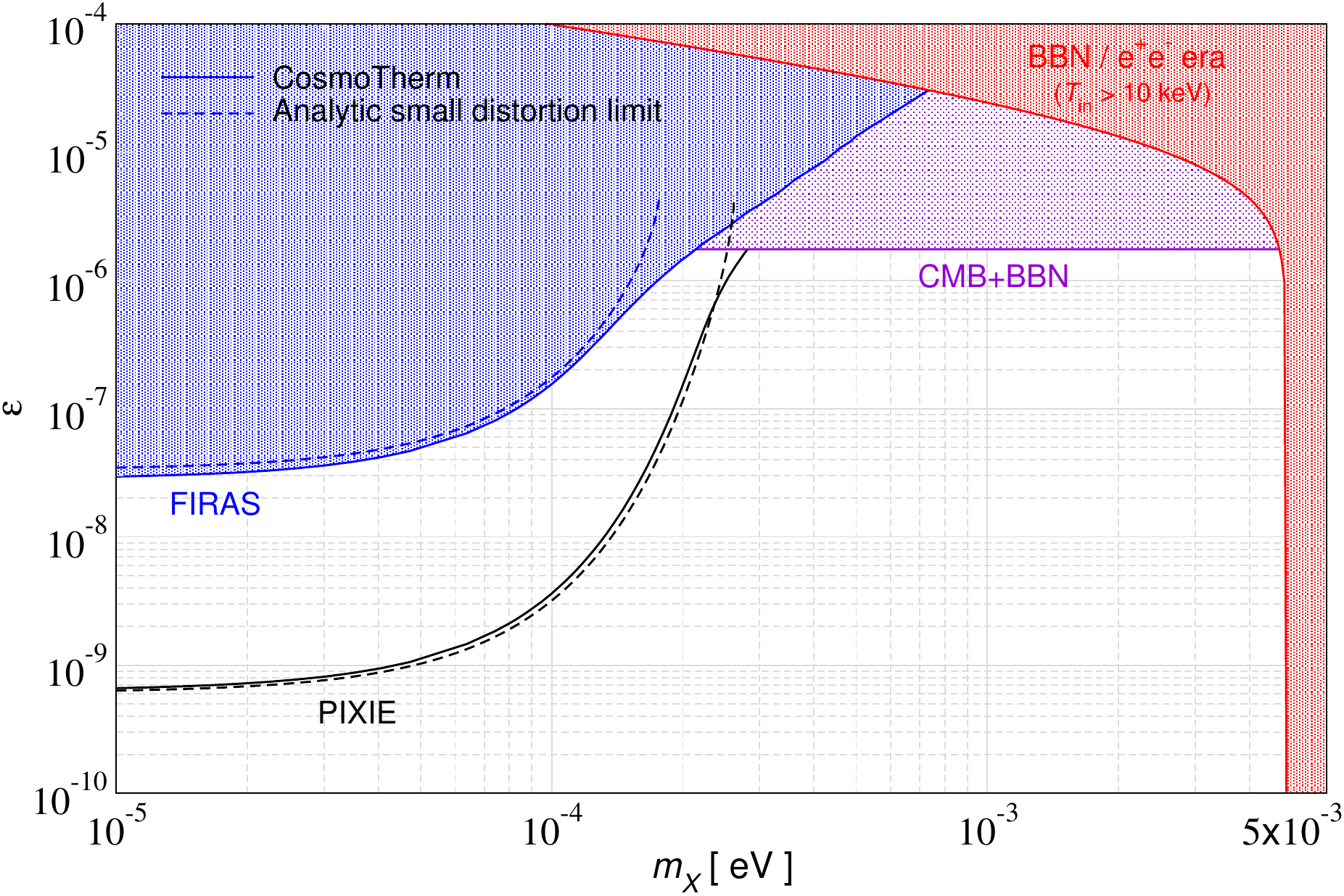}
\caption{CMB spectral distortion limits from \COBEF and \PIXIE in comparison to the CMB+BBN limit on $N_{\rm eff}$ (all at 95\% c.l.).}
\label{fig:Limits_distortion_BBN}
\end{figure}
%
\subsection{Comparing to constraints from $\Delta N_{\rm eff}$}
At $\mdp > 10^{-4}\,\eV$, additional constraints can be obtained by considering changes to 
$N_{\rm eff}$ \citep[e.g.,][]{Jaeckel2008, McDermott2020}. Strictly speaking, simple treatments are only possible for scenarios with photon to dark photon conversions at (initial) temperatures $\lesssim 10\,\keV$ (i.e., post-BBN / e$^+$e$^-$ annihilation). For conversions during BBN a complete treatment including the co-evolution of the distorted photon field and time-dependent changes to the nuclear rates is necessary but beyond the current state-of-the-art \citep[see][for review and latest computations]{Steigman2007, Pitrou2020BBNQED}. For a pre-BBN treatment, the decoupling of neutrinos from the thermal bath has to be distinguished. To avoid these complications, we therefore restrict ourselves to constraints on post-BBN conversions, meaning masses $\mdp\lesssim \pot{5}{-3}\,\eV$.

For the constraints, only the total energy conversion, $\Delta \rho/\rho\big|_{\rm con}$, matters as one does not need to know the precise photon distribution.
The CMB energy density changes to $\rho_{\rm CMB}=\rho_{\rm CMB, in}(1+\epsilon_\rho)$. 
%
Before the conversion, one has the initial energy density $\rho_{\rm CMB, in}$ at a temperature $\Tin>\TCMB=T_0(1+\zcon)$. We can also write
the total radiation energy density as
\begin{align}
\rho_{\rm rad, in}
=\rho_{\rm CMB, in}\left[1+\kappa_\nu \,N_{\rm eff}\right],
\end{align}
with $\kappa_\nu=(7/8)\,(4/11)^{4/3}\approx 0.2271$ at redshifts right after the BBN era ended. After the photon conversion, we then have
\begin{align}
\rho_{\rm rad}
&=\rho_{\rm CMB, in}(1+\epsilon_\rho)+\rho_{\rm CMB, in}\,\kappa_\nu \,N_{\rm eff}-\rho_{\rm CMB, in}\epsilon_\rho
\nonumber\\
&=\rho_{\rm CMB}\left[1+\kappa_\nu \,\left(\frac{N_{\rm eff}}{1+\epsilon_\rho}-\frac{\epsilon_\rho}{\kappa_\nu(1+\epsilon_\rho)}\right)\right],
\end{align}
where the last quantity in parenthesis is the apparent value of $N_{\rm eff}$ after the conversion \citep[see also][]{Jaeckel2008}.

To obtain the constraint, we have to compare the value of $N^{\rm BBN}_{\rm eff}$ during the BBN era with an independent measurement from the CMB, which primarily relies on changes to the matter-radiation equality and hence early ISW effect. Using $N^{\rm BBN}_{\rm eff}=3.046$ and comparing to $N^{\rm CMB}_{\rm eff}=2.99\pm0.34$ from \Planck+BAO \citep{Planck2018params} one can then obtain $|\epsilon_\rho|\lesssim 0.1$ (95\% c.l.) using a simple $\chi^2$ approach. This implies $\epsilon<\pot{1.8}{-6}$ (95\% c.l.) at $\mdp \lesssim \pot{5}{-3}\,\eV$.
We compare this limit with the CMB distortion constraints in Fig.~\ref{fig:Limits_distortion_BBN}. The parameter space ruled out by CMB spectral distortions is increased over the simple small distortion limits and the $N_{\rm eff}$ constraint is superseded at $\mdp\lesssim \pot{2}{-4}\,\eV$ by \COBEF, while \PIXIE could push this to $\mdp\lesssim \pot{3}{-4}\,\eV$.
We avoid the region where the end of BBN overlaps with the conversion (red domain in Fig.~\ref{fig:Limits_distortion_BBN}).

At this point it is important to mention several caveats. First of all, we have assumed that the dark photon background stays relativistic until $z\simeq 10^{3}$. The removed CMB photons have an average energy of 
\begin{align}
\frac{\left<\omega\right>}{\Tin} \approx \frac{\int x^3 \Delta n \id x}{\int x^2 \Delta n \id x}=
\frac{\int \frac{x^3}{\expf{x}-1} \left[1-\expf{-\gammacon/x}\right] \id x}{\int \frac{x^2}{\expf{x}-1} \left[1-\expf{-\gammacon/x}\right] \id x}.
\end{align}
With $\gammacon \approx 0.31$ and $\Tin \approx \TCMB$ (these are valid for $\epsilon=\pot{1.8}{-6}$), we then have $\left<\omega\right>\approx 1.7 \,\TCMB(\zcon)$. For the relevant mass-range this means $\left<\omega\right>\approx 0.8\,\keV-8\,\keV\gg \mdp$. Consequently, these dark photons should still be mostly relativistic at $z\simeq 10^3$ but could turn partially non-relativistic below $z\lesssim 10$ due to redshifting. A more detailed estimate is presented in Appendix~\ref{sec:app-A} and yields a very similar conclusion.

Secondly, we mention that for a consistent treatment of the $N_{\rm eff}$ limit derived above one should also include the change of the baryon to photon ratio $\eta_{\rm b}=N_{\rm b}/N_\gamma$, which during BBN would be lower than the present value \citep{Steigman2007}. This changes the precise value of the expected $N^{\rm BBN}_{\rm eff}$, and also reduces the value for the helium mass fraction $Y_{\rm p}$. For a consistent analysis, one should therefore analyze the \Planck data with a modified value of $Y_{\rm p}$. However, $\epsilon_\rho=-0.1$ means $\Delta \Tin/\TCMB\approx 0.027$. This implies $\Delta \eta_{\rm b}/\eta_{\rm b}\approx -0.08$ and hence a change of $Y_{\rm p}$ by less than\footnote{We estimated this change using Eq.~(8) of \citet{Steigman2007}.} 0.0002, which can be safely neglected.

Another constraint can also be derived by comparing different measurements of $\eta_{\rm b}$ \citep[see][]{Simha2008, Jaeckel2008}; however, we do not go into further details here since this comes out weaker. Future $N_{\rm eff}$ constraints promise uncertainties at the level $\Delta N_{\rm eff}\approx 0.05$ \citep{SOWP2018}. This should improve the constraints on $\epsilon_\rho$ by a factor of $\simeq 3-4$. Even in this case, CMB spectrometers provide tighter constraints for $\mdp \lesssim \pot{2}{-4}\,\eV$.

\vspace{-3mm}
\section{Discussion and Conclusions} \label{sec:level5}
The spectral distortion measurements from \COBEF have celebrated their 30 year anniversary. Nevertheless, this legacy data set still provides the most stringent constraints on many new physics models. Spectral distortion calculations have now become sufficiently accurate to reach the full potential of the \COBEF data and in this work we translate these measurements into limits on photon to dark photon conversion processes in the post-BBN era.

We have shown that previous estimates of these limits did not predict the correct sign and amplitude of the distortion signal due to the omission of entropy extraction. This tightens the distortion bounds by roughly a factor of 1.5 (Fig.~\ref{fig:Limits_distortion}). We furthermore treated the large distortion limit of the distortion evolution allowing us to go beyond the small distortion approximation. Correspondingly, the distortion constraints on dark photon masses in the range $\mdp\simeq 10^{-4}\,\eV-10^{-3}\,\eV$ were significantly underestimated in previous works (Fig.~\ref{fig:Limits_distortion_BBN}).
We also comment that existing spectral distortion constraints on photon to axion conversion \citep[e.g.,][]{Tashiro2013} suffer from similar issues with regards to entropy extraction. We plan to investigate this problem next.

Our analysis also identified a generic distortion shape for the dark photon signal, Eq.~\eqref{eq:Dn_dp_late}. At CMB frequencies, the photon conversion process sources a temperature shift {\it and} a $\mu$-type distortion, almost independent of the conversion redshift. This opens the possibility to study anisotropic distortions from dark photon conversion processes using recently developed methods \citep[e.g.,][]{Chluba2023I, Chluba2023II, Kite2023III}. A $\mu-T$ correlation is expected and can in principle be constrained with \Planck \citep[e.g.,][]{Rotti2022, Bianchini2022}, {\it Litebird} \citep{Remazeilles2022muT}, CMB-S4 \citep{Zegeye2023} and the SKA \citep{Zegeye2024}, as highlighted in the context of primordial non-Gaussianity \citep[e.g.,][]{Pajer2012, Ganc2012, Biagetti2013}. 
A detection of the related $\mu-T$ correlation would complement the recent limits presented for masses $\mdp\simeq 10^{-12}\,\eV$ based on anisotropic conversion at low redshifts \citep{McCarthy2024patchy}, allowing to extend these to higher masses using CMB imaging. However, we leave a more detailed discussion to future work.

Finally we mention that this is the first example of a process that can source (average) $\mu$-type distortions even after the $\mu$-era. The distortions are not exactly degenerate, but our study implies that it becomes harder to identify the primordial origin of a distortion signal. Nevertheless, this makes dark photon distortions an exciting new target for future spectrometers like BISOU or \PIXIE that is not degenerate with the late time $y$-type distortion.

\section*{Acknowledgements}
JC would like to thank Junwu Huang and Cristina Mondino for insightful comments and great hospitality during his visit to Perimeter Institute.
JC was supported by the ERC Consolidator Grant {\it CMBSPEC} (No.~725456) and by the Royal Society as a Royal Society University Research Fellow at the University of Manchester, UK (No.~URF/R/191023).
BC would like to acknowledge support from NSERC postdoctoral and Banting fellowships.
MCJ is supported by the Natural Sciences and Engineering Research Council of Canada through a Discovery grant. Research at Perimeter Institute is supported in part by the Government of Canada through the Department of Innovation, Science and Economic Development Canada and by the Province of Ontario through the Ministry of Research, Innovation and Science.

\section{Data availability}
The data underlying in this article are available in this article and can further be made available on request.

{\small
\vspace{-3mm}
\bibliographystyle{mn2e}
\bibliography{Lit, lit_I}

\begin{thebibliography}{82}
\expandafter\ifx\csname natexlab\endcsname\relax\def\natexlab#1{#1}\fi

\bibitem[{{Abitbol} {et~al}\mbox{.}(2017){Abitbol}, {Chluba}, {Hill}, \&
  {Johnson}}]{abitbol_pixie}
{Abitbol} M.~H., {Chluba} J., {Hill} J.~C., {Johnson} B.~R., 2017, \mnras

\bibitem[{{Acharya} \& {Chluba}(2022)}]{Acharya2022large}
{Acharya} S.~K., {Chluba} J., 2022, \mnras, 515, 5775

\bibitem[{{Acharya} {et~al}\mbox{.}(2021){Acharya}, {Chluba}, \&
  {Sarkar}}]{Acharya2021}
{Acharya} S.~K., {Chluba} J., {Sarkar} A., 2021, \mnras, 507, 2052

\bibitem[{{Acharya} {et~al}\mbox{.}(2023){Acharya}, {Cyr}, \&
  {Chluba}}]{Acharya2023SPH}
{Acharya} S.~K., {Cyr} B., {Chluba} J., 2023, \mnras, 523, 1908

\bibitem[{{Alonso-{\'A}lvarez} {et~al}\mbox{.}(2020){Alonso-{\'A}lvarez},
  {Gupta}, {Jaeckel}, \& {Spannowsky}}]{Alonso2020Wondrous}
{Alonso-{\'A}lvarez} G., {Gupta} R.~S., {Jaeckel} J., {Spannowsky} M., 2020,
  \jcap, 2020, 052

\bibitem[{{Andr{\'e}} {et~al}\mbox{.}(2014){Andr{\'e}}, {Baccigalupi},
  {Banday}, {Barbosa}, {Barreiro}, {Bartlett}, {Bartolo}, {Battistelli},
  {Battye}, {Bendo}, {Beno{\#523}t}, {Bernard}, {Bersanelli}, {B{\'e}thermin},
  {Bielewicz}, {Bonaldi}, {Bouchet}, {Boulanger}, {Brand}, {Bucher},
  {Burigana}, {Cai}, {Camus}, {Casas}, {Casasola}, {Castex}, {Challinor},
  {Chluba}, {Chon}, {Colafrancesco}, {Comis}, {Cuttaia}, {D'Alessandro}, {Da
  Silva}, {Davis}, {de Avillez}, {de Bernardis}, {de Petris}, {de Rosa}, {de
  Zotti}, {Delabrouille}, {D{\'e}sert}, {Dickinson}, {Diego}, {Dunkley},
  {En{\ss}lin}, {Errard}, {Falgarone}, {Ferreira}, {Ferri{\`e}re}, {Finelli},
  {Fletcher}, {Fosalba}, {Fuller}, {Galli}, {Ganga}, {Garc{\'{\i}}a-Bellido},
  {Ghribi}, {Giard}, {Giraud-H{\'e}raud}, {Gonzalez-Nuevo}, {Grainge},
  {Gruppuso}, {Hall}, {Hamilton}, {Haverkorn}, {Hernandez-Monteagudo},
  {Herranz}, {Jackson}, {Jaffe}, {Khatri}, {Kunz}, {Lamagna}, {Lattanzi},
  {Leahy}, {Lesgourgues}, {Liguori}, {Liuzzo}, {Lopez-Caniego}, {Macias-Perez},
  {Maffei}, {Maino}, {Mangilli}, {Martinez-Gonzalez}, {Martins}, {Masi},
  {Massardi}, {Matarrese}, {Melchiorri}, {Melin}, {Mennella}, {Mignano},
  {Miville-Desch{\^e}nes}, {Monfardini}, {Murphy}, {Naselsky}, {Nati},
  {Natoli}, {Negrello}, {Noviello}, {O'Sullivan}, {Paci}, {Pagano}, {Paladino},
  {Palanque-Delabrouille}, {Paoletti}, {Peiris}, {Perrotta}, {Piacentini},
  {Piat}, {Piccirillo}, {Pisano}, {Polenta}, {Pollo}, {Ponthieu},
  {Remazeilles}, {Ricciardi}, {Roman}, {Rosset}, {Rubino-Martin}, {Salatino},
  {Schillaci}, {Shellard}, {Silk}, {Starobinsky}, {Stompor}, {Sunyaev},
  {Tartari}, {Terenzi}, {Toffolatti}, {Tomasi}, {Trappe}, {Tristram},
  {Trombetti}, {Tucci}, {Van de Weijgaert}, {Van Tent}, {Verde}, {Vielva},
  {Wandelt}, {Watson}, \& {Withington}}]{PRISM2013WPII}
{Andr{\'e}} P. {et~al.}, 2014, \jcap, 2, 6

\bibitem[{Arias {et~al}\mbox{.}(2012)Arias, Cadamuro, Goodsell, Jaeckel,
  Redondo, \& Ringwald}]{Arias2012}
Arias P., Cadamuro D., Goodsell M., Jaeckel J., Redondo J., Ringwald A., 2012,
  JCAP, 06, 013

\bibitem[{{Biagetti} {et~al}\mbox{.}(2013){Biagetti}, {Perrier}, {Riotto}, \&
  {Desjacques}}]{Biagetti2013}
{Biagetti} M., {Perrier} H., {Riotto} A., {Desjacques} V., 2013, \prd, 87,
  063521

\bibitem[{{Bianchini} \& {Fabbian}(2022)}]{Bianchini2022}
{Bianchini} F., {Fabbian} G., 2022, \prd, 106, 063527

\bibitem[{{Bolliet} {et~al}\mbox{.}(2020){Bolliet}, {Chluba}, \&
  {Battye}}]{Bolliet2020PI}
{Bolliet} B., {Chluba} J., {Battye} R., 2020, arXiv e-prints, arXiv:2012.07292

\bibitem[{Bolliet {et~al}\mbox{.}(2021)Bolliet, Chluba, \&
  Battye}]{Bolliet2020}
Bolliet B., Chluba J., Battye R., 2021, \mnras, 507, 3148

\bibitem[{Brahma {et~al}\mbox{.}(2023)Brahma, Berlin, \& Schutz}]{Brahma2023}
Brahma N., Berlin A., Schutz K., 2023, Phys. Rev. D, 108, 095045

\bibitem[{{Burigana} {et~al}\mbox{.}(1991){Burigana}, {Danese}, \& {de
  Zotti}}]{Burigana1991}
{Burigana} C., {Danese} L., {de Zotti} G., 1991, \aap, 246, 49

\bibitem[{{Cabass} {et~al}\mbox{.}(2016){Cabass}, {Melchiorri}, \&
  {Pajer}}]{Cabass2016}
{Cabass} G., {Melchiorri} A., {Pajer} E., 2016, \prd, 93, 083515

\bibitem[{Caputo {et~al}\mbox{.}(2020)Caputo, Liu, Mishra-Sharma, \&
  Ruderman}]{Caputo2020}
Caputo A., Liu H., Mishra-Sharma S., Ruderman J.~T., 2020, Phys. Rev. Lett.,
  125, 221303

\bibitem[{Caputo {et~al}\mbox{.}(2021)Caputo, Millar, O'Hare, \&
  Vitagliano}]{Caputo2021}
Caputo A., Millar A.~J., O'Hare C. A.~J., Vitagliano E., 2021, Phys. Rev. D,
  104, 095029

\bibitem[{{Chluba}(2005)}]{Chluba2005Thesis}
{Chluba} J., 2005, PhD thesis, Ludwig-Maximilians University of Munich, Germany

\bibitem[{{Chluba}(2013{\natexlab{a}})}]{Chluba2013fore}
{Chluba} J., 2013{\natexlab{a}}, \mnras, 436, 2232

\bibitem[{{Chluba}(2013{\natexlab{b}})}]{Chluba2013Green}
{Chluba} J., 2013{\natexlab{b}}, \mnras, 434, 352

\bibitem[{{Chluba}(2014)}]{Chluba2014}
{Chluba} J., 2014, \mnras, 440, 2544

\bibitem[{{Chluba}(2015)}]{Chluba2015GreensII}
{Chluba} J., 2015, \mnras, 454, 4182

\bibitem[{{Chluba}(2016)}]{Chluba2016}
{Chluba} J., 2016, \mnras, 460, 227

\bibitem[{{Chluba} {et~al}\mbox{.}(2021){Chluba}, {Abitbol}, {Aghanim},
  {Ali-Ha{\"\i}moud}, {Alvarez}, {Basu}, {Bolliet}, {Burigana}, {de Bernardis},
  {Delabrouille}, {Dimastrogiovanni}, {Finelli}, {Fixsen}, {Hart},
  {Hern{\'a}ndez-Monteagudo}, {Hill}, {Kogut}, {Kohri}, {Lesgourgues},
  {Maffei}, {Mather}, {Mukherjee}, {Patil}, {Ravenni}, {Remazeilles}, {Rotti},
  {Rubi{\~n}o-Martin}, {Silk}, {Sunyaev}, \& {Switzer}}]{Chluba2021Voyage}
{Chluba} J. {et~al.}, 2021, Experimental Astronomy, 51, 1515

\bibitem[{{Chluba} {et~al}\mbox{.}(2012{\natexlab{a}}){Chluba}, {Erickcek}, \&
  {Ben-Dayan}}]{Chluba2012inflaton}
{Chluba} J., {Erickcek} A.~L., {Ben-Dayan} I., 2012{\natexlab{a}}, \apj, 758,
  76

\bibitem[{{Chluba} \& {Jeong}(2014)}]{Chluba2013PCA}
{Chluba} J., {Jeong} D., 2014, \mnras, 438, 2065

\bibitem[{{Chluba} {et~al}\mbox{.}(2012{\natexlab{b}}){Chluba}, {Khatri}, \&
  {Sunyaev}}]{Chluba2012}
{Chluba} J., {Khatri} R., {Sunyaev} R.~A., 2012{\natexlab{b}}, \mnras, 425,
  1129

\bibitem[{{Chluba} {et~al}\mbox{.}(2023{\natexlab{a}}){Chluba}, {Kite}, \&
  {Ravenni}}]{Chluba2023I}
{Chluba} J., {Kite} T., {Ravenni} A., 2023{\natexlab{a}}, \jcap, 2023, 026

\bibitem[{{Chluba} {et~al}\mbox{.}(2019){Chluba}, {Kogut}, {Patil}, {Abitbol},
  {Aghanim}, {Ali-Ha{\i}{\ensuremath{\ddot{}}}moud}, {Amin}, {Aumont},
  {Bartolo}, {Basu}, {Battistelli}, {Battye}, {Baumann}, {Ben-Dayan},
  {Bolliet}, {Bond}, {Bouchet}, {Burgess}, {Burigana}, {Byrnes}, {Cabass},
  {Chuss}, {Clesse}, {Cole}, {Dai}, {de Bernardis}, {Delabrouille},
  {Desjacques}, {de Zotti}, {Diacoumis}, {Dimastrogiovanni}, {Di Valentino},
  {Dunkley}, {Durrer}, {Dvorkin}, {Ellis}, {Eriksen}, {Fasiello}, {Fixsen},
  {Finelli}, {Flauger}, {Galli}, {Garcia-Bellido}, {Gervasi}, {Gluscevic},
  {Grin}, {Hart}, {Hern{\'a}ndez-Monteagudo}, {Hill}, {Jeong}, {Johnson},
  {Lagache}, {Lee}, {Lewis}, {Liguori}, {Kamionkowski}, {Khatri}, {Kohri},
  {Komatsu}, {Kunze}, {Mangilli}, {Masi}, {Mather}, {Matarrese},
  {Miville-Desch{\^e}nes}, {Montaruli}, {M{\"u}nchmeyer}, {Mukherjee},
  {Nakama}, {Nati}, {Ota}, {Page}, {Pajer}, {Poulin}, {Ravenni}, {Reichardt},
  {Remazeilles}, {Rotti}, {Rubi{\~n}o-Martin}, {Sarkar}, {Sarkar}, {Savini},
  {Scott}, {Serpico}, {Silk}, {Souradeep}, {Spergel}, {Starobinsky},
  {Subrahmanyan}, {Sunyaev}, {Switzer}, {Tartari}, {Tashiro}, {Thakur},
  {Trombetti}, {Wallisch}, {Wandelt}, {Wehus}, {Wollack}, {Zaldarriaga}, \&
  {Zannoni}}]{Chluba2019}
{Chluba} J. {et~al.}, 2019, \baas, 51, 184

\bibitem[{{Chluba} {et~al}\mbox{.}(2020{\natexlab{a}}){Chluba}, {Ravenni}, \&
  {Acharya}}]{Chluba2020large}
{Chluba} J., {Ravenni} A., {Acharya} S.~K., 2020{\natexlab{a}}, \mnras, 498,
  959

\bibitem[{{Chluba} {et~al}\mbox{.}(2020{\natexlab{b}}){Chluba}, {Ravenni}, \&
  {Bolliet}}]{Chluba2020BRpack}
{Chluba} J., {Ravenni} A., {Bolliet} B., 2020{\natexlab{b}}, \mnras, 492, 177

\bibitem[{{Chluba} {et~al}\mbox{.}(2023{\natexlab{b}}){Chluba}, {Ravenni}, \&
  {Kite}}]{Chluba2023II}
{Chluba} J., {Ravenni} A., {Kite} T., 2023{\natexlab{b}}, \jcap, 2023, 027

\bibitem[{{Chluba} {et~al}\mbox{.}(2007){Chluba}, {Sazonov}, \&
  {Sunyaev}}]{Chluba2007a}
{Chluba} J., {Sazonov} S.~Y., {Sunyaev} R.~A., 2007, \aap, 468, 785

\bibitem[{{Chluba} \& {Sunyaev}(2012)}]{Chluba2011therm}
{Chluba} J., {Sunyaev} R.~A., 2012, \mnras, 419, 1294

\bibitem[{{Cyr} {et~al}\mbox{.}(2024){Cyr}, {Acharya}, \&
  {Chluba}}]{Cyr2024SPH}
{Cyr} B., {Acharya} S.~K., {Chluba} J., 2024, arXiv e-prints, arXiv:2404.11743

\bibitem[{Cyr {et~al}\mbox{.}(2023)Cyr, Chluba, \& Acharya}]{Cyr2023}
Cyr B., Chluba J., Acharya S.~K., 2023, \mnras, 525, 2632

\bibitem[{Cyr {et~al}\mbox{.}(2024)Cyr, Kite, Chluba, Hill, Jeong, Acharya,
  Bolliet, \& Patil}]{Cyr2023GW}
Cyr B., Kite T., Chluba J., Hill J.~C., Jeong D., Acharya S.~K., Bolliet B.,
  Patil S.~P., 2024, \mnras, 528, 883

\bibitem[{{Fixsen}(2003)}]{Fixsen2003}
{Fixsen} D.~J., 2003, \apjl, 594, L67

\bibitem[{{Fixsen} {et~al}\mbox{.}(1996){Fixsen}, {Cheng}, {Gales}, {Mather},
  {Shafer}, \& {Wright}}]{Fixsen1996}
{Fixsen} D.~J., {Cheng} E.~S., {Gales} J.~M., {Mather} J.~C., {Shafer} R.~A.,
  {Wright} E.~L., 1996, \apj, 473, 576

\bibitem[{{Ganc} \& {Komatsu}(2012)}]{Ganc2012}
{Ganc} J., {Komatsu} E., 2012, \prd, 86, 023518

\bibitem[{{Hill} {et~al}\mbox{.}(2015){Hill}, {Battaglia}, {Chluba}, {Ferraro},
  {Schaan}, \& {Spergel}}]{Hill2015}
{Hill} J.~C., {Battaglia} N., {Chluba} J., {Ferraro} S., {Schaan} E., {Spergel}
  D.~N., 2015, Physical Review Letters, 115, 261301

\bibitem[{Holdom(1986)}]{Holdom1985}
Holdom B., 1986, Phys. Lett. B, 166, 196

\bibitem[{{Hu} \& {Silk}(1993)}]{Hu1993}
{Hu} W., {Silk} J., 1993, \prd, 48, 485

\bibitem[{{Jackson}(1998)}]{Jackson}
{Jackson} J.~D., 1998, {Classical Electrodynamics, 3rd Edition}. Wiley-VCH

\bibitem[{{Jaeckel} {et~al}\mbox{.}(2008){Jaeckel}, {Redondo}, \&
  {Ringwald}}]{Jaeckel2008}
{Jaeckel} J., {Redondo} J., {Ringwald} A., 2008, \prl, 101, 131801

\bibitem[{{Khatri} \& {Sunyaev}(2012)}]{Khatri2012b}
{Khatri} R., {Sunyaev} R.~A., 2012, \jcap, 6, 38

\bibitem[{{Khatri} {et~al}\mbox{.}(2012){Khatri}, {Sunyaev}, \&
  {Chluba}}]{Khatri2011BE}
{Khatri} R., {Sunyaev} R.~A., {Chluba} J., 2012, \aap, 540, A124

\bibitem[{{Kite} {et~al}\mbox{.}(2023){Kite}, {Ravenni}, \&
  {Chluba}}]{Kite2023III}
{Kite} T., {Ravenni} A., {Chluba} J., 2023, \jcap, 2023, 028

\bibitem[{{Kogut} {et~al}\mbox{.}(2016){Kogut}, {Chluba}, {Fixsen}, {Meyer}, \&
  {Spergel}}]{Kogut2016SPIE}
{Kogut} A., {Chluba} J., {Fixsen} D.~J., {Meyer} S., {Spergel} D., 2016, in
  Proc.SPIE, Vol. 9904, SPIE Conference Series, p. 99040W

\bibitem[{{Kogut} {et~al}\mbox{.}(2011){Kogut}, {Fixsen}, {Chuss}, {Dotson},
  {Dwek}, {Halpern}, {Hinshaw}, {Meyer}, {Moseley}, {Seiffert}, {Spergel}, \&
  {Wollack}}]{Kogut2011PIXIE}
{Kogut} A. {et~al.}, 2011, \jcap, 7, 25

\bibitem[{{Kogut} {et~al}\mbox{.}(2024){Kogut}, {Switzer}, {Fixsen}, {Aghanim},
  {Chluba}, {Chuss}, {Delabrouille}, {Dvorkin}, {Hensley}, {Hill}, {Maffei},
  {Pullen}, {Rotti}, {Sabyr}, {Thiele}, {Wollack}, \& {Zelko}}]{KogutPIXIE2024}
{Kogut} A. {et~al.}, 2024, arXiv e-prints, arXiv:2405.20403

\bibitem[{{Lucca} {et~al}\mbox{.}(2020){Lucca}, {Sch{\"o}neberg}, {Hooper},
  {Lesgourgues}, \& {Chluba}}]{Lucca2020}
{Lucca} M., {Sch{\"o}neberg} N., {Hooper} D.~C., {Lesgourgues} J., {Chluba} J.,
  2020, \jcap, 2020, 026

\bibitem[{{Maffei} {et~al}\mbox{.}(2021){Maffei}, {Abitbol}, {Aghanim},
  {Aumont}, {Battistelli}, {Chluba}, {Coulon}, {De Bernardis}, {Douspis},
  {Grain}, {Gervasoni}, {Hill}, {Kogut}, {Masi}, {Matsumura}, {Sullivan},
  {Pagano}, {Pisano}, {Remazeilles}, {Ritacco}, {Rotti}, {Sauvage}, {Savini},
  {Stever}, {Tartari}, {Thiele}, \& {Trappe}}]{BISOU}
{Maffei} B. {et~al.}, 2021, arXiv e-prints, arXiv:2111.00246

\bibitem[{{Masi} {et~al}\mbox{.}(2021){Masi}, {Battistelli}, {de Bernardis},
  {Coppolecchia}, {Columbro}, {D'Alessandro}, {De Petris}, {Lamagna},
  {Marchitelli}, {Mele}, {Paiella}, {Piacentini}, {Pisano}, {Bersanelli},
  {Franceschet}, {Manzan}, {Mennella}, {Realini}, {Cibella}, {Martini},
  {Pettinari}, {Coppi}, {Gervasi}, {Limonta}, {Zannoni}, {Piccirillo}, \&
  {Tucker}}]{Masi2021}
{Masi} S. {et~al.}, 2021, arXiv e-prints, arXiv:2110.12254

\bibitem[{{McCarthy} {et~al}\mbox{.}(2024){McCarthy}, {Pirvu}, {Hill}, {Huang},
  {Johnson}, \& {Rogers}}]{McCarthy2024patchy}
{McCarthy} F., {Pirvu} D., {Hill} J.~C., {Huang} J., {Johnson} M.~C., {Rogers}
  K.~K., 2024, arXiv e-prints, arXiv:2406.02546

\bibitem[{McDermott \& Witte(2020)}]{McDermott2019}
McDermott S.~D., Witte S.~J., 2020, Phys. Rev. D, 101, 063030

\bibitem[{{McDermott} \& {Witte}(2020)}]{McDermott2020}
{McDermott} S.~D., {Witte} S.~J., 2020, \prd, 101, 063030

\bibitem[{Mirizzi {et~al}\mbox{.}(2009{\natexlab{a}})Mirizzi, Redondo, \&
  Sigl}]{Mirizzi2009b}
Mirizzi A., Redondo J., Sigl G., 2009{\natexlab{a}}, JCAP, 08, 001

\bibitem[{Mirizzi {et~al}\mbox{.}(2009{\natexlab{b}})Mirizzi, Redondo, \&
  Sigl}]{Mirizzi2009a}
Mirizzi A., Redondo J., Sigl G., 2009{\natexlab{b}}, JCAP, 03, 026

\bibitem[{Okun(1982)}]{Okun1982}
Okun L.~B., 1982, Sov. Phys. JETP, 56, 502

\bibitem[{{Pajer} \& {Zaldarriaga}(2012)}]{Pajer2012}
{Pajer} E., {Zaldarriaga} M., 2012, Physical Review Letters, 109, 021302

\bibitem[{{Pitrou} \& {Pospelov}(2020)}]{Pitrou2020BBNQED}
{Pitrou} C., {Pospelov} M., 2020, \prc, 102, 015803

\bibitem[{{Planck Collaboration} {et~al}\mbox{.}(2020){Planck Collaboration},
  {Aghanim}, {Akrami}, {Ashdown}, {Aumont}, {Baccigalupi}, {Ballardini},
  {Banday}, {Barreiro}, {Bartolo}, {Basak}, {Battye}, {Benabed}, {Bernard},
  {Bersanelli}, {Bielewicz}, {Bock}, {Bond}, {Borrill}, {Bouchet}, {Boulanger},
  {Bucher}, {Burigana}, {Butler}, {Calabrese}, {Cardoso}, {Carron},
  {Challinor}, {Chiang}, {Chluba}, {Colombo}, {Combet}, {Contreras}, {Crill},
  {Cuttaia}, {de Bernardis}, {de Zotti}, {Delabrouille}, {Delouis}, {Di
  Valentino}, {Diego}, {Dor{\'e}}, {Douspis}, {Ducout}, {Dupac}, {Dusini},
  {Efstathiou}, {Elsner}, {En{\ss}lin}, {Eriksen}, {Fantaye}, {Farhang},
  {Fergusson}, {Fernandez-Cobos}, {Finelli}, {Forastieri}, {Frailis},
  {Fraisse}, {Franceschi}, {Frolov}, {Galeotta}, {Galli}, {Ganga},
  {G{\'e}nova-Santos}, {Gerbino}, {Ghosh}, {Gonz{\'a}lez-Nuevo}, {G{\'o}rski},
  {Gratton}, {Gruppuso}, {Gudmundsson}, {Hamann}, {Handley}, {Hansen},
  {Herranz}, {Hildebrandt}, {Hivon}, {Huang}, {Jaffe}, {Jones}, {Karakci},
  {Keih{\"a}nen}, {Keskitalo}, {Kiiveri}, {Kim}, {Kisner}, {Knox},
  {Krachmalnicoff}, {Kunz}, {Kurki-Suonio}, {Lagache}, {Lamarre}, {Lasenby},
  {Lattanzi}, {Lawrence}, {Le Jeune}, {Lemos}, {Lesgourgues}, {Levrier},
  {Lewis}, {Liguori}, {Lilje}, {Lilley}, {Lindholm}, {L{\'o}pez-Caniego},
  {Lubin}, {Ma}, {Mac{\'\i}as-P{\'e}rez}, {Maggio}, {Maino}, {Mandolesi},
  {Mangilli}, {Marcos-Caballero}, {Maris}, {Martin}, {Martinelli},
  {Mart{\'\i}nez-Gonz{\'a}lez}, {Matarrese}, {Mauri}, {McEwen}, {Meinhold},
  {Melchiorri}, {Mennella}, {Migliaccio}, {Millea}, {Mitra},
  {Miville-Desch{\^e}nes}, {Molinari}, {Montier}, {Morgante}, {Moss}, {Natoli},
  {N{\o}rgaard-Nielsen}, {Pagano}, {Paoletti}, {Partridge}, {Patanchon},
  {Peiris}, {Perrotta}, {Pettorino}, {Piacentini}, {Polastri}, {Polenta},
  {Puget}, {Rachen}, {Reinecke}, {Remazeilles}, {Renzi}, {Rocha}, {Rosset},
  {Roudier}, {Rubi{\~n}o-Mart{\'\i}n}, {Ruiz-Granados}, {Salvati}, {Sandri},
  {Savelainen}, {Scott}, {Shellard}, {Sirignano}, {Sirri}, {Spencer},
  {Sunyaev}, {Suur-Uski}, {Tauber}, {Tavagnacco}, {Tenti}, {Toffolatti},
  {Tomasi}, {Trombetti}, {Valenziano}, {Valiviita}, {Van Tent}, {Vibert},
  {Vielva}, {Villa}, {Vittorio}, {Wandelt}, {Wehus}, {White}, {White},
  {Zacchei}, \& {Zonca}}]{Planck2018params}
{Planck Collaboration} {et~al.}, 2020, \aap, 641, A6

\bibitem[{{Ravenni} \& {Chluba}(2020)}]{Ravenni2020DC}
{Ravenni} A., {Chluba} J., 2020, \jcap, 2020, 025

\bibitem[{Redondo(2008)}]{Redondo2008a}
Redondo J., 2008, JCAP, 07, 008

\bibitem[{{Refregier} {et~al}\mbox{.}(2000){Refregier}, {Komatsu}, {Spergel},
  \& {Pen}}]{Refregier2000}
{Refregier} A., {Komatsu} E., {Spergel} D.~N., {Pen} U.-L., 2000, \prd, 61,
  123001

\bibitem[{{Remazeilles} {et~al}\mbox{.}(2022){Remazeilles}, {Ravenni}, \&
  {Chluba}}]{Remazeilles2022muT}
{Remazeilles} M., {Ravenni} A., {Chluba} J., 2022, \mnras, 512, 455

\bibitem[{{Rotti} \& {Chluba}(2021)}]{Rotti2021}
{Rotti} A., {Chluba} J., 2021, \mnras, 500, 976

\bibitem[{{Rotti} {et~al}\mbox{.}(2022){Rotti}, {Ravenni}, \&
  {Chluba}}]{Rotti2022}
{Rotti} A., {Ravenni} A., {Chluba} J., 2022, \mnras, 515, 5847

\bibitem[{{Rubi{\~n}o Mart{\'\i}n} {et~al}\mbox{.}(2020){Rubi{\~n}o
  Mart{\'\i}n}, {Alonso Arias}, {Hoyland}, {Aguiar-Gonz{\'a}lez}, {De
  Miguel-Hern{\'a}ndez}, {G{\'e}nova-Santos}, {Gomez-Re{\~n}asco}, {Guidi},
  {Fern{\'a}ndez-Izquierdo}, {Fern{\'a}ndez-Torreiro}, {Fuerte-Rodriguez},
  {Hernandez-Monteagudo}, {L{\'o}pez-Caraballo}, {Perez-de-Taoro}, {Peel},
  {Rebolo}, {Zamora-Jimenez}, {Gonz{\'a}lez-Carretero}, {Colodro-Conde},
  {P{\'e}rez-Lemus}, {Toledo-Moreo}, {P{\'e}rez-Liz{\'a}n}, {Cuttaia},
  {Terenzi}, {Franceschet}, {Realini}, {Chluba}, {Murga-Llano}, \&
  {Sanquirce-Garcia}}]{Jose2020TMS}
{Rubi{\~n}o Mart{\'\i}n} J.~A. {et~al.}, 2020, in Society of Photo-Optical
  Instrumentation Engineers (SPIE) Conference Series, Vol. 11453, Society of
  Photo-Optical Instrumentation Engineers (SPIE) Conference Series, p. 114530T

\bibitem[{{Sarkar} {et~al}\mbox{.}(2019){Sarkar}, {Chluba}, \&
  {Lee}}]{Sarkar2019}
{Sarkar} A., {Chluba} J., {Lee} E., 2019, \mnras, 490, 3705

\bibitem[{{Sathyanarayana Rao} {et~al}\mbox{.}(2015){Sathyanarayana Rao},
  {Subrahmanyan}, {Udaya Shankar}, \& {Chluba}}]{Mayuri2015}
{Sathyanarayana Rao} M., {Subrahmanyan} R., {Udaya Shankar} N., {Chluba} J.,
  2015, \apj, 810, 3

\bibitem[{{Simha} \& {Steigman}(2008)}]{Simha2008}
{Simha} V., {Steigman} G., 2008, \jcap, 6, 16

\bibitem[{{Steigman}(2007)}]{Steigman2007}
{Steigman} G., 2007, Annual Review of Nuclear and Particle Science, 57, 463

\bibitem[{{Sunyaev} \& {Khatri}(2013)}]{Sunyaev2013}
{Sunyaev} R.~A., {Khatri} R., 2013, IJMPD, 22, 30014

\bibitem[{{Sunyaev} \& {Zeldovich}(1970{\natexlab{a}})}]{Sunyaev1970mu}
{Sunyaev} R.~A., {Zeldovich} Y.~B., 1970{\natexlab{a}}, \apss, 7, 20

\bibitem[{{Sunyaev} \& {Zeldovich}(1970{\natexlab{b}})}]{Sunyaev1970SPEC}
{Sunyaev} R.~A., {Zeldovich} Y.~B., 1970{\natexlab{b}}, Comments on
  Astrophysics and Space Physics, 2, 66

\bibitem[{{Tashiro}(2014)}]{Tashiro2014}
{Tashiro} H., 2014, Prog. of Theo. and Exp. Physics, 2014, 060000

\bibitem[{{Tashiro} {et~al}\mbox{.}(2013){Tashiro}, {Silk}, \&
  {Marsh}}]{Tashiro2013}
{Tashiro} H., {Silk} J., {Marsh} D.~J.~E., 2013, \prd, 88, 125024

\bibitem[{{The Simons Observatory Collaboration} {et~al}\mbox{.}(2018){The
  Simons Observatory Collaboration}, {Ade}, {Aguirre}, {Ahmed}, {Aiola}, {Ali},
  {Alonso}, {Alvarez}, {Arnold}, {Ashton}, \& et~al.}]{SOWP2018}
{The Simons Observatory Collaboration} {et~al.}, 2018, ArXiv:1808.07445

\bibitem[{{Zegeye} {et~al}\mbox{.}(2023){Zegeye}, {Bianchini}, {Bond},
  {Chluba}, {Crawford}, {Fabbian}, {Gluscevic}, {Grin}, {Hill}, {Meerburg},
  {Orlando}, {Partridge}, {Reichardt}, {Remazeilles}, {Scott}, {Wollack}, \&
  {CMB-S4 Collaboration}}]{Zegeye2023}
{Zegeye} D. {et~al.}, 2023, \prd, 108, 103536

\bibitem[{{Zegeye} {et~al}\mbox{.}(2024){Zegeye}, {Crawford}, {Chluba},
  {Remazeilles}, \& {Grainge}}]{Zegeye2024}
{Zegeye} D., {Crawford} T., {Chluba} J., {Remazeilles} M., {Grainge} K., 2024,
  arXiv e-prints, arXiv:2406.04326

\bibitem[{{Zeldovich} \& {Sunyaev}(1969)}]{Zeldovich1969}
{Zeldovich} Y.~B., {Sunyaev} R.~A., 1969, \apss, 4, 301

\end{thebibliography}
}

\appendix

\section{Relativistic Fraction} \label{sec:app-A}
At the conversion redshift, an $x$-dependent fraction of CMB photons convert into the dark sector. The lowest energy photons to convert are those at the plasma frequency ($E_{\gamma} = \omega_{\rm pl}$), which in principle become zero-momentum dark photons. On the other hand, photons with frequency $\omega \gtrsim \sqrt{2}\omega_{\rm pl} = \sqrt{2} m_{\rm dp}$ will convert and produce a relativistic population\footnote{We define a relativistic dark photon as one in which $p_{\rm d}/m_{\rm d} \geq 1$.} of dark photons. As the conversion of a relativistic species (the CMB photons) into a non-relativistic one can alter the expansion history, and thus $N_{\rm eff}$, it is instructive to determine the fraction of relativistic vs non-relativistic dark photons at some redshift $z \leq z_{\rm con}$.

\begin{figure}
\includegraphics[width=\columnwidth]{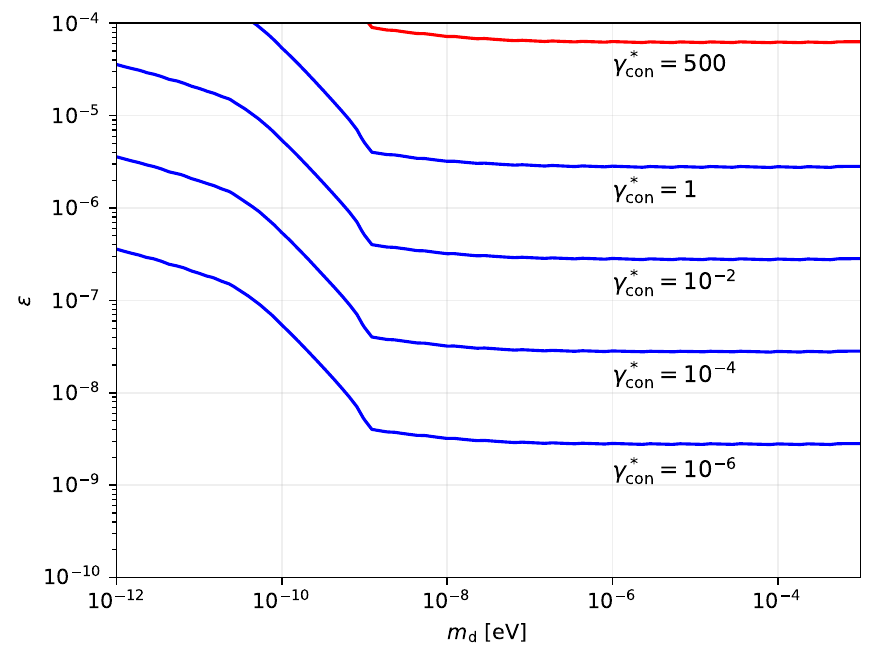}
\caption{Contours of constant $\gamma^*_{\rm con}$ over the dark photon parameter space we consider. The approximate solutions for $\epsilon_{\rho}$ and $T_{\rm in}$ given in Eqs.~\eqref{eq:rho_approx} and \eqref{eq:Tin_approx}, are valid below the red contour, with maximal error (up to $15$\%) around $\gamma^*_{\rm con} = 1$.}
\label{fig:gamma_contour}
\end{figure}

At the time of conversion, the energy density that is transferred into the dark sector is given by the energy extracted from the CMB,
\begin{align} 
    \rho_{\rm dp} = \frac{T_{\rm in}^4}{\pi^2} \int_{\omega_{\rm pl}/T_{\rm in}}^{\infty} \id x \, x^3 \bigg(1- {\rm e}^{-\gamma^*_{\rm con}/x} \bigg) \, n_{\rm bb}(x).
\end{align}
The lower bound ensures conversions below the plasma frequency don't take place, and it is understood that here $x = \omega/\Tin$. Contours of constant $\gamma^*_{\rm con}$ can be found in Fig.~\ref{fig:gamma_contour} for the parameter space we consider. The amplitude of $\gamma^*_{\rm con}$ determines the size of the CMB temperature shift when a conversion takes place. For $\gamma^*_{\rm con} \ll 1$, we have $\Tin \simeq T_{\rm CMB}$ as discussed in the main text. 

After conversion, dark photons are produced with some spectrum of initial state momenta given by $p(z_{\rm con})$. After production, that momentum will redshift such that such that $p(z < z_{\rm con}) = p(z_{\rm con})\left(\frac{1+z}{1+z_{\rm con}}\right)$. From this we find that photons with
\begin{align}
    \omega_{\rm rel} \geq m_{\rm dp} \left( 1 + \left(\frac{1+z_{\rm con}}{1+z} \right)^2\right)^{1/2}
\end{align}
at the conversion redshift will lead to relativistic dark photons at some later redshift $z$. We can then define the non-relativistic fraction as
\begin{align}
    f_{\rm NR}(z) = \frac{\int^{x_{\rm rel}(z)}_{x_{\rm pl}} \id x \frac{x^2}{{\rm e}^{x}-1}\left(1-{\rm e}^{-\gamma^*_{\rm con}/x}\right)}{\int^{\infty}_{x_{\rm pl}} \id x \frac{x^2}{{\rm e}^{x}-1} \left(1-{\rm e}^{-\gamma^*_{\rm con}/x}\right)},
\end{align}
and relativistic fraction $f_{\rm R} = 1 - f_{\rm NR}$. Note that $x_{\rm pl} = \omega_{\rm pl}/T_{\rm in} = m_{\rm dp}/T_{\rm in}$ at $z_{\rm con}$.

The value of $\gamma^*_{\rm con}$ encodes the fraction of the initial CMB that is converted. As the peak of the CMB lies around $x \simeq 1$, values of $\gamma^*_{\rm con}$ larger or smaller than this value will convert more or less of the bulk of the CMB. However, for a fixed $m_{\rm dp}$, larger values of $\gamma^*_{\rm con}$ will actually translate into smaller fractions of non-relativistic photons at a later time since more high energy dark photons are created. 

\begin{figure}
\includegraphics[width=\columnwidth]{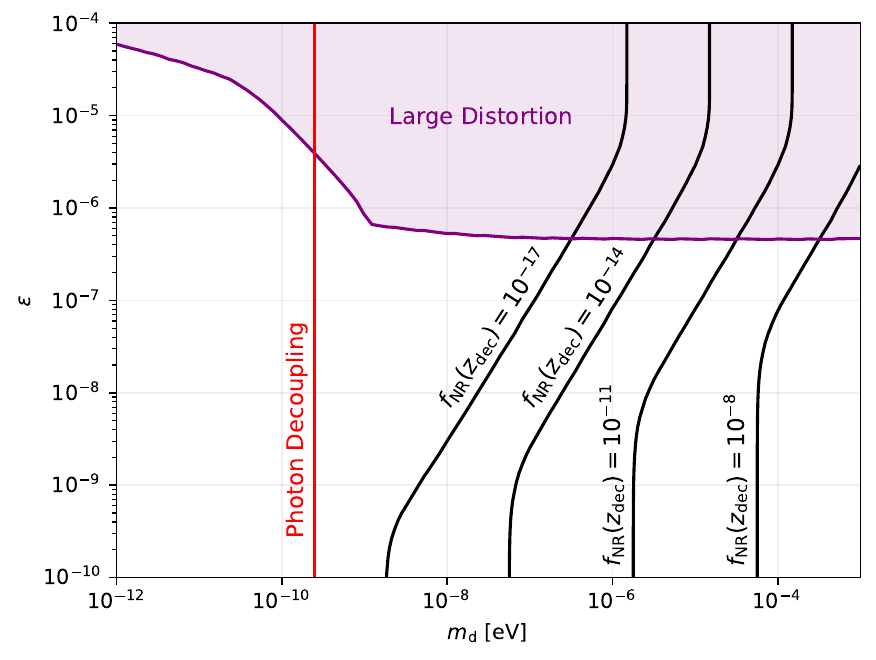}
\caption{Contours of constant $f_{\rm NR}$ evaluated at photon decoupling. While instructive, for this particular model the time evolution of $f_{\rm NR}$ between $z_{\rm con}$ and decoupling is negligibly small.}
\label{fig:f_NR_contour}
\end{figure}

Constraints on $N_{\rm eff}$ can be discerned both at BBN and recombination, and so it is useful to understand how this spectrum of non-thermal dark photons evolves. In principle one would expect an evolution in $N_{\rm eff}$ for these types of models. In Fig.~\ref{fig:f_NR_contour}, we show contours of constant $f_{\rm NR}$, evaluated at the photon decoupling redshift. 

In the large $\epsilon$ limit, the contours become vertical due to the fact that the majority of the initial CMB has converted, and thus the non-relativistic fraction freezes in. In the opposite ($\epsilon \ll 1$) limit, we find that $\gamma^*_{\rm con}/x \ll 1$, hence the fraction once again becomes independent of $\epsilon$. The contours slope to the right in each instance because for a given $m_{\rm d}$ (and hence a given $z_{\rm con}$) the \textit{total} energy density in non-relativistic dark photons remains constant, while the total energy density $\rho_{\rm dp}$ continues to increase with increasing $\epsilon$. The ratio therefore continues to shrink. For all cases considered here, the redshift evolution of $f_{\rm NR}$ between BBN and decoupling is far below the uncertainty on either measurement of $N_{\rm eff}$.

%
%

%
%

\end{document}